\documentclass[useAMS,usenatbib]{mn2e}
\usepackage{natbib}
\usepackage{epsfig}
\usepackage{graphicx,psfrag,amssymb}
\usepackage{bm}
\usepackage{aas_macros}

\newcommand{\arcdeg}{\ensuremath{^{\circ}}}
\title[Understanding roAp stars]{On the understanding of pulsations in the atmosphere of roAp stars: phase diversity and false nodes}
\author[J. C. Sousa and M. S. Cunha]{J. C. Sousa $^{1,2}$ \thanks{E-mail:
jsousa@astro.up.pt} and M. S. Cunha $^{1,2}$\\
$^1$ Centro de Astrof\'{\i}sica da Universidade do Porto\\
$^2$ Faculdade de Ci\^encias da Universidade do Porto, Rua das Estrelas, 4150-762, Porto, Portugal
}

\begin{document}

\date{Accepted 1988 December 15. Received 1988 December 14; in original form 1988 October 11}

\pagerange{\pageref{firstpage}--\pageref{lastpage}} \pubyear{2002}

\maketitle

\label{firstpage}

\begin{abstract}
Studies based on high-resolution spectroscopic data of rapidly oscillating Ap stars show a surprising diversity of pulsation behavior in the atmospheric layers, pointing, in particular, to the co-existence of running and standing waves. The correct interpretation of these data requires a careful modelling of pulsations in these magnetic stars. In light of this, in this work we present a theoretical analysis of pulsations in roAp stars, taking into account the direct influence of the magnetic field. We derive approximate analytical solutions for the displacement components parallel and perpendicular to the direction of the magnetic field, that are appropriate to the outermost layer. From these, we determine the expression for the theoretical radial velocity for an observer at a general position, and compute the corresponding pulsation amplitude and phase as function of height in the atmosphere. We show that the integral for the radial velocity has contributions from three different types of wave solutions, namely, running waves, evanescent waves, and standing waves of nearly constant amplitude. We then consider a number of case studies to illustrate the origin of the different pulsational behaviour that is found in the observations. Concerning pulsation amplitude, we find that it generally increases with atmospheric height. Pulsation phase, however, shows a diversity of behaviours, including phases that are constant, increasing, or decreasing with atmospheric height. Finally, we show that there are situations in which the pulsation amplitude goes through a zero, accompanied by a phase jumps of $\pi$, and argue that such behaviour does not correspond to a pulsation node in the outermost layers of the star, but rather to a visual effect, resulting from the observers inability to resolve the stellar surface.
\end{abstract}


\section{Introduction}
			The rapidly oscillating Ap (roAp) stars are found among the coolest subgroup of chemically peculiar Ap stars, which is located in the main sequence part of the classical instability strip. They pulsate with periods typically within the range from 5 to 21 minutes, (e.g. \citet{2005MNRAS.358L...6K}), have oscillations with amplitudes between 0.5 and 5 km\,s$^{-1}$ in velocity, and, as in other Ap stars, have strong, large scale magnetic fields, with typical intensities of a few kG, although in some stars the magnetic field strength can be higher than 20 kG,(e.g, \citet{2009MNRAS.396.1018H}). Moreover, they have about two solar masses \citep{1990ARA&A..28..607K}, and temperatures that range from about 6400 to 8100 K (\citet{2009CoAst.159...61K}).\ 

		The present number of known roAp stars is more than 40. Due to their characteristics, with roAp stars we have the unique opportunity to observe the interaction of acoustic modes with strong large-scale, magnetic fields. The roAp stars have been observed photometrically since their discovery by \citet{1982MNRAS.200..807K} and until recently the knowledge about their acoustic oscillations was essentially based on high-speed, ground-based, photometric observations.  However, in the past few years numerous exciting observational results have been published, both as a result of the acquisition of space-based photometry \citep[e.g.][]{2008A&A...480..223G,2009MNRAS.396.1189B,2010arXiv1006.4013B} and as a result of the analysis of high-resolution spectroscopic data \citep[e.g.][]{2006ESASP.624E..33K,2006MNRAS.370.1274K,2007A&A...473..907R,2007CoAst.150...81S}. Such high-resolution spectroscopic data hold unique information about the structure and dynamics of the peculiar atmospheres of roAp stars, and have revealed a surprising diversity in the pulsation behavior of different lines in the roAp spectra.\   

The general picture that emerges from the analysis of time-series of high-resolution spectroscopic data of roAp stars is that pulsational variability is seen predominantly in lines of rare-earth ions, especially those of Pr and Nd, which are strong and numerous in the roAp spectra and are formed in the higher layers of the atmosphere.  On the other hand, lines of light and iron-peak elements, enhanced in the lower atmospheric layers, often do not show pulsation variability within the observational detection limit for most roAp stars \citep[e.g.][]{2002A&A...384..545R}.  Thus, generally pulsations are found to be weak or non-detectable in the lower atmosphere, while often reach amplitudes of several km\,s$^{-1}$ higher in the atmosphere.
  
Another interesting feature that can be inferred from spectroscopic observations of roAp stars is the presence of significant shifts in pulsation phase when comparing radial velocities derived from lines of different rare-earth elements, or even from different lines of the same element \citep[e.g.][]{2001A&A...374..615K}.  Likewise, phase shifts are seen when performing depth-in-line analysis of the H$\alpha$ and Nd III 6145\,\AA\ lines, \citep[e.g.][]{2006ESASP.624E..33K}. Along with the phase shifts, depth variations of pulsation amplitude are also inferred, which in the generality of the cases point to an increase of amplitude with atmospheric height. Moreover, in some stars the radial velocity amplitudes measured in lines of the rare-earth ion Pr III and in the core of the H$\alpha$ line, are seen to vary on a time scale of a few pulsation cycles \citep{2006MNRAS.370.1274K}.

The phase behaviour derived from spectroscopic studies of roAp stars is usually interpreted as resulting from the presence of running waves, or standing waves, in the atmosphere, depending, respectively, on whether, or not, the phase shows a variation with depth. In particular, it was found by \citet{2007CoAst.150...81S} and \citet{2007A&A...473..907R}, that for stars with pulsation frequencies below the acoustic cut-off frequency, the pulsations seem to have standing wave character in the deeper layers and then, in the outer layers, behave like running waves. On the other hand, for stars which have pulsation frequency close to, or higher than, the cut-off frequency, the authors found that the pulsations behave like running waves from the deepest layers.	
 
 One important and interesting feature that can be seen in spectroscopic data of roAp stars is the existence of pulsation zeros in the amplitude diagram accompanied by phase jumps in the phase diagram. This was interpreted as the presence of a pulsation node at some depth in the atmosphere. The first alleged node found in a roAp star was detected by \citet{1998MNRAS.295...33B} in the atmosphere of $\alpha$ Cir. More recently, in the case of the star 10Aql, \citet{2008MNRAS.tmp..334E} found that the pulsation phase derived from the lines of Tb III and Dy III differ by $180$ \arcdeg. According to the authors this could be an indication of a radial node between the line-forming layers of these elements. Moreover, they found that the line bisectors for strong NdIII line profiles show significant changes of phase, or even phase jumps, as in the case of the line NdIII at 5102.41 \AA. For this line, the authors found a phase shift of $180$\arcdeg and at the same depth, an amplitude close to zero, which suggests the existence of a radial node. In addition, \citet{2003MNRAS.345..781M} studied the radial velocity variations in the roAp star 33 Librae. The authors found that the NdIII lines pulsate nearly in anti-phase with those of NdII, a feature that the authors attribute to the presence of a pulsation node in the atmosphere of this star. The pulsational behaviour of the Nd ii and Nd iii lines in 33 Lib has also been discussed by \citet{2005MNRAS.358L...6K}. They found that the line depth versus pulsation amplitude supported the hypothesis of a node between the line-forming layers of these two ions. \citet{2007A&A...473..907R} found similar results for this star, confirming a phase jump between the NdIII and NdII lines, and the amplitude decreasing towards zero at the same atmospheric height.\

The results of the our theoretical analysis will be presented in two separate papers, as detailed in sec.~\ref{sec:parameters}. In sec.~\ref{sec:theory} we discuss the physics underlying our model, along with the assumptions made. In 
sec.~\ref{RV} we derive the expression for the theoretical radial velocity and describe a Toy Model that will be useful for the interpretation of the results. The results of the analysis of six case studies are presented in sec.~\ref{results}, followed by a general discussion in sec.~\ref{conclusions}.

\section{Parameter space}
\label{sec:parameters}
Given an underlying equilibrium stellar model, the general problem that we set out to study
%, namely the form of the amplitudes and phases of pulsational radial velocity as derived from the analysis of time-series of high resolution spectroscopy,
 depends on a number of input parameters. Some of these are intrinsic to the star, namely: the magnetic field intensity and topology, characterized by the vector $\bm{B}$; the unperturbed (i.e., in the absence of a magnetic field) oscillation mode, characterized by its radial order, $n$, angular degree, $l$, and the azimuthal order $m$; the location of the chemical elements whose spectral lines are used in the derivation of the radial velocity, characterized by an atmospheric depth (or region in depth), as well as by longitudinal and latitudinal limits. Moreover, some parameters depend on the position of the observer, such as: the inclination angle between the magnetic axis and the direction of the observer $\delta$ (at a given phase of rotation); the inclination angle between the latter and the rotation axis, $i$.

In our analysis we will keep some of the above fixed. In particular, we will only consider magnetic fields of a dipolar topology. We will not consider the effect of rotation on the dynamics and, consequently, will assume that the magnetic and pulsation axis are aligned. This is generally a good approximation for roAp stars, except possibly in cases when the manetic field is rather low (below $\approx $1 kG) \citep{2002A&A...391..235B}. Moreover, the unperturbed pulsation modes to be considered are only those axisymmetric about the magnetic axis, thus, characterized by $m=0$ in a spherical harmonic decomposition about that axis of symmetry.  Finally, we will consider only a fixed phase of rotation, so that our problem will have no dependence on the inclination angle $i$.  We note that the relaxation of most of the above conditions is relatively straightforward, although in some cases will lead to significant additional work. 

The results of our study will be presented in two papers. The present paper will deal with the underlying mathematical analysis and the in-depth investigation of a number of test cases -- the{\it case studies} discussed in section~\ref{results}. For that, we will restrict the parameter space further, by considering only modes of degree $l=1$ and the case in which the pulsation and magnetic axis is along the line-of-sight (i.e., $\delta=0$). The latter is justified by the fact that in this case the spherical coordinate system associated with the observer coincides with the one used to solve the pulsation problem in the star. As will become clear in sec.~\ref{toy model}, that is a necessary condition for the in-depth study that we aim at in the present paper.

In a second paper we will present the conclusions of a systematic search of the parameter space, including also variation of mode degree and of the position of the observer. Additionally, in that paper we will compare our results with those derived by simulations of line profiles, as well as with general trends of pulsation phase and amplitude found in the observations.

\section{Pulsations in the presence of a magnetic field}
\label{sec:theory}
	
%	As mentioned before, high-resolution spectroscopic studies of roAp stars have revealed a surprising diversity in the pulsation behavior of different lines in the roAp spectra. With the purpose of understanding these observations we have carried out a theoretical analysis of the pulsations in the outermost layers of these stars.   
The study of pulsations in the presence of a magnetic field has been undertaken by a number of authors in the past, adopting different approximations, or mathematical approaches \citep{1996ApJ...458..338D,2000A&A...356..218B,2000MNRAS.319.1020C,2004MNRAS.350..485S,2005MNRAS.360.1022S,2006MNRAS.365..153C}.

\begin{figure}
\begin{center}
\includegraphics[width=20pc]{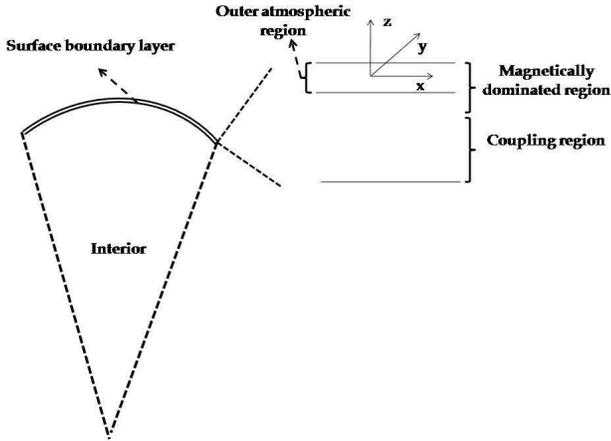}\hspace{4pc}
\caption{Schematic representation of the star, vertical cut.}
\label{fig.camadas_estrela}
\end{center}
\end{figure}

The magnetic field has a direct influence on the oscillations only in the region where the Lorentz forces are comparable to or larger than the gas pressure gradient, that is to say, in the surface layers of the stars. Therefore, we follow the approach of previous studies, and divide the star into two regions, a region where the magnetic pressure is greater than or comparable to the gas pressure, the surface boundary layer, and a region where the gas pressure is much larger than the magnetic pressure, the interior, defined to be the region between the center of the star and the boundary layer. In the interior the dynamics is not directly affected by the magnetic field and the oscillations there will be modified only as a consequence of the change in the conditions of the surface layers.

  The surface boundary layer itself can be divided into two regions, one where the magnetic pressure is comparable to the gas pressure, that we shall call the coupling region, and one where the magnetic pressure is much larger than the gas pressure, which is the magnetically dominated region. 
  
  Concerning the oscillations, the studies referred to above have shown that in the coupling region the magnetic and acoustic wave components are coupled into a magnetoacoustic wave. Moreover, the authors have shown that in the interior of the star, where the gas pressure is much larger than the magnetic pressure, the magnetoacustic wave decouples into a fast (acoustic in nature) and a slow (magnetic in nature) component, the latter being an inwardly propagating wave. Likewise, higher in the atmosphere, where the magnetic pressure is much larger than the gas pressure, the magnetocaoustic wave decouples into acoustic and magnetic components, the latter being a standing wave with almost constant amplitude \citep{2008MNRAS.tmp..337S}.

In Fig. \ref{fig.camadas_estrela}, we show a schematic representation of a vertical cut in the star. On the left hand side we show the interior and the surface boundary layer and on the right hand side we show a zoom of the surface boundary layer in which the coupling and magnetically dominated regions are depicted. Moreover, the figure shows also the outer atmospheric region, which we define as a subsection of the magnetically dominated region in which the plasma is assumed to be approximately isothermal. In this study we will concentrate in the outer atmospheric region, and will take as input the velocity field at the base of that region which, in turn, is obtained from the study of the star as a whole, following the approach of \citet{2006MNRAS.365..153C}.\

\subsection{Surface boundary layer}

In our calculations we follow \citet{2000MNRAS.319.1020C} and assume a dipolar magnetic field, $\bm{B}$, of polar magnitude $B_p$ which is force free, and hence does not influence the equilibrium state. Therefore, in this layer the equilibrium state is governed by the following system of equations,

\begin{equation}
\label{eq: 2.7}
\frac{\partial\rho_0}{\partial t}=0 ,
\end{equation}  

\begin{equation}
\label{eq: 2.8}
\frac{\partial p_0}{\partial t}=0 ,
\end{equation}  

\begin{equation}
\label{eq: 2.9}
\nabla p_0= \rho_0\bm{g}_0,
\end{equation}  
where $p$ is the pressure, $\rho$ is the density, $\bm{g}$ is the gravitational field, $t$ is the time, and the subscript $0$ denotes the equilibrium quantities. The equilibrium gravitational field, $\bm{g}_0$, can be written as the gradient of the gravitational potential, $\Phi$,

\begin {equation}
\label{eq: 2.10v}
\bm{g}_0= -\nabla \Phi,   
\end{equation}
where $\Phi$ satisfies Poisson's equation,

\begin {equation}
\label{eq: 2.10p}
\nabla^2\Phi = 4\pi G\rho_0,   
\end{equation}
where $G$ is the gravitational constant. 

In the limit of perfect conductivity, adiabatic motions associated with a velocity, $\bm{v}$, are governed by the following system of magneto-hydrodynamic equations,\

\begin {equation}
\label{eq: 2.10}
\frac{\partial \bm{B}}{\partial t}=\nabla\times(\bm{v}\times \bm{B}),   
\end{equation}

\begin {equation}
\label{eq: 2.11}
\frac{d \rho}{d t} + \rho\nabla.\bm{v} = 0 ,	
\end{equation}

\begin {equation}
\label{eq: 2.12}
\frac{d p}{d t} = \frac{\gamma p}{\rho}\frac{d\rho}{dt} , 
\end{equation}

\begin {equation}
\label{eq: 2.13}
\rho\frac{d \bm{v}}{d t} = -\nabla p +\bm{j}\times \bm{B} +\rho \bm{g} ,
\end{equation}
where $d$ stands for total derivatives, $\partial$ stands for partial derivatives, $\gamma$ is the first adiabatic exponent and $\bm{j}$ is the current density. \

Assuming small adiabatic perturbations of the equilibrium state, such that,

\begin{eqnarray}
q=q_0+q_1 \nonumber,
\end{eqnarray}
for any scalar or vector component, $q$, and neglecting the terms of higher order in the perturbations, the two previous systems of equations can be rearranged to give a new system of equations which govern linear adiabatic pulsations in the Cowling approximation, namely,    

\begin{equation}
\label{eq: 2.14}
\rho_0\frac{\partial^2 \bm{\bar{\xi}}}{\partial t^2}= -\nabla p_1+{\rho_1}{\bm g_0}+\frac{1}{\mu_0}\left(\nabla\times \bm{B_1}\right) \times \bm{B_0},
\end{equation}

\begin{equation}
\label{eq: 2.15}
p_1=-\bm{\bar{\xi}}.\nabla p_0-\gamma p_0\nabla.\bm{\bar{\xi}},
\end{equation}

\begin{equation}
\label{eq: 2.16}
\rho_1=-\bm{\bar{\xi}}.\nabla \rho_0-\rho_0\nabla.\bm{\bar{\xi}},
\end{equation}

\begin{equation}
\label{eq: 2.17}
\bm{B_1}=\nabla\times(\bm{\bar{\xi}}\times \bm{B_0}),
\end{equation}
where $\mu_0$ is the magnetic permeability and $\bm{\bar{\xi}}$ is the vector displacement, defined by the relation $\bm{v_1}$=$\partial \bm{\bar{\xi}}/\partial t$, and where we used the fact that the magnetic field in the equilibrium state is irrotational.\

As expected, the magnetic term in equation (\ref{eq: 2.14}) is perpendicular to the direction of $\bm{B}_0$ and, thus, within the approximations considered, there is no magnetic component of the restoring force along the direction of the unperturbed magnetic field. 
To proceed we follow the assumptions adopted in previous studies and consider, at each latitude, a local plane-parallel layer with locally constant $\bm{B}_0=\left[B_x,0,B_z\right]$ and $\bm{g_0}$, and neglect $\partial \bm{B}_0/\partial x$ and $\partial \bm{B}_0/\partial z$, but allow $\bm{B}$ to change with latitude. In the above, we considered a local coordinate system (\textit{x},\textit{y},\textit{z}), with \textit{x} and \textit{y} pointing along the latitudinal and longitudinal directions, respectively. Moreover, \textit{z} is taken to be zero at some fiducial radius (that shall be identified later) located in the outer atmospheric region, and increasing outwardly.  

\subsubsection{Components of the displacement}

Let us first begin with the projections of equation (\ref{eq: 2.14}) in the directions parallel and perpendicular to the magnetic field. For that we will define $\bar{\xi}_{||}$ and $\bar{\xi}_{\bot}$ such that $\bm{\bar{\xi}}={\bar{\xi}}_{||}\bm{\hat{e}}_{||}+ {\bar{\xi}}_{\bot}\bm{\hat{e}}_{\bot}$, with $\bm{\hat{e}}_{||}=\bm{B_0}/|\bm{B_0}|$, and $\bm{\hat{e}}_{\bot}=(\bm{B_0}\times\bm{\hat{e}}_{y})/|\bm{B_0}|$, the unit vectors in the direction parallel and perpendicular to $\bm{B}_0$, respectively, and $\bm{\hat{e}}_x$ and $\bm{\hat{e}}_y$ the unit vectors in the direction of the \textit{x} and \textit{y} axes, of the adopted Cartesian coordinate system, respectively. We note that due to the axisymmetry of both the magnetic field and the oscillation modes to be considered, the problem reduces to fourth-order (two second order coupled differential equations), and the only projections that need to be considered in the analysis lie in the ($\bm{\hat{e}}_x$, $\bm{\hat{e}}_z$) plane.

 Since the magnetic term in equation (\ref{eq: 2.14}) is zero along the direction of the magnetic field, when combined with equations (\ref{eq: 2.15}) and (\ref{eq: 2.16}), the projection of equation (\ref{eq: 2.14}) in this direction becomes,

\begin{equation}
\label{eq:parallel}
\bm{\hat{e}}_{||}.\left[\rho_0\frac{\partial^2 \bm{\bar{\xi}}}{\partial t^2}\right]=\gamma\nabla_{||}(p_0\nabla.\bm{\bar{\xi}})+\rho_0 g (\bm{\hat{e}}_x.\nabla){\bar{\xi}}_{\bot},
\end{equation} 
where $\nabla_{||}$ is the directional derivative in the direction parallel to the magnetic field which is defined as $\nabla_{||}=\bm{\hat{e}}_{||}.\nabla$, and $\bm{g_0}=g \bm{\hat{e}}_{z}$.

Next, we assume that all horizontal derivatives of the displacement, and its components, are much smaller than the corresponding vertical derivatives. For low degree, high radial order modes in the absence of a magnetic field this is clearly a good approximation. In the presence of a magnetic field there are additional dependencies of the eigenfunctions on the latitude, but these occur on a scale of $1/R$, where $R$ is the radius of the star, and thus introduce local horizontal derivatives that are much smaller than the vertical derivatives in the atmosphere. Thus, this approximation remains good in general, in the presence of a magnetic field.\
Note, however, that this approximation may fail, at particular latitudes, for particular frequencies, when the phase and/or amplitude variations of the displacement with latitude are very sharp \citep[e.g.][]{2000MNRAS.319.1020C}. However, since the observer sees only an integral of the eigenfunctions over the stellar disk, that failure is likely not to have a strong impact on the results, unless the eigenfunction is trapped within a given latitudinal region, in which case the approach considered here will not be adequate.

So, neglecting the horizontal derivatives of the displacement when compared with the vertical derivatives, we get, after some algebra, the following expression for the projection of equation (\ref{eq: 2.14}) in the direction along the magnetic field, 

\begin{eqnarray}
\frac{\rho_0 \left|\bm{B}_0\right|^2}{\gamma p_0 B_z^2}\frac{\partial^2 {\bar{\xi}}_{||}}{\partial t^2}=\frac{\partial^2 {\bar{\xi}}_{||}}{\partial z^2}+\frac{B_x}{B_z}\frac{\partial^2 {\bar{\xi}}_{\bot}}{\partial z^2}+\nonumber
\end{eqnarray} 

\begin{equation}
\ \ \ \ \ \ \ \ \ \ \ \ \ \ \ \ \ \ \ \ \  \frac{1}{p_0}\frac{dp_0}{dz}\left(\frac{\partial {\bar{\xi}}_{||}}{\partial z}+\frac{B_x}{B_z}\frac{\partial {\bar{\xi}}_{\bot}}{\partial z}\right).
\label{eq:parallel_disp}
\end{equation} 

Let us now consider the projection of equation (\ref{eq: 2.14}) in the direction perpendicular to the magnetic field. Taking the scalar product with $\bm{\hat{e}}_{\bot}$ we get,

\begin{eqnarray}
\bm{\hat{e}}_{\bot}.\left[\rho_0\frac{\partial^2 \bm{\bar{\xi}}}{\partial t^2}\right]=\bm{\hat{e}}_{\bot}.[-\nabla p_1+ \rho_1 \bm{g}_0] + \nonumber
\end{eqnarray}

\begin{equation}
\ \ \ \ \ \ \ \ \ \ \ \ \ \ \ \ \ \ \ \ \frac{1}{\mu_0}\left(\left|\bm{B}_0\right|\nabla_{||}{B}_{1\bot}-\left|\bm{B}_0\right|\nabla_{\bot}{B}_{1||}\right),
\label{eq:perpendicular_versor}
\end{equation}
where $\nabla_{\bot}$ is the directional derivative in the direction perpendicular to the magnetic field, defined as $\nabla_{\bot}=\bm{\hat{e}}_{\bot}.\nabla$, and ${B}_{1\bot}$ and ${B}_{1||}$ are, respectively, the components of $\bm{B}_1$ along and perpendicular to the magnetic field direction.\

In equation (\ref{eq:perpendicular_versor}), the first term on the right hand side corresponds to the restoring force that would act in the absence of the magnetic field while the second term corresponds to the magnetic response. After developing both terms and combining with equations (\ref{eq: 2.15}), (\ref{eq: 2.16}) and (\ref{eq: 2.17}), equation (\ref{eq:perpendicular_versor}) becomes,

\begin{eqnarray}
\rho_0\frac{\partial^2 {\bar{\xi}}_{\bot}}{\partial t^2}=\frac{\left|\bm{B}_0\right|^2}{\mu_0}\nabla^2 {\bar{\xi}}_{\bot}+\gamma\nabla_{\bot}(p_0\nabla.\bm{\bar{\xi}}) + \nonumber
\end{eqnarray} 

\begin{equation}
\ \ \ \ \ \ \ \ \ \ \ \ \ \ \ \rho_0 g\left(\frac{B_z}{\left|\bm{B}_0\right|}\nabla_{\bot}{\bar{\xi}}_{||}-\frac{B_x}{\left|\bm{B}_0\right|}\nabla_{||}{\bar{\xi}}_{||}\right).
\label{perpendicular_1}
\end{equation} 

Furthermore, similarly to what we have done for the projection in the direction along the magnetic field, we neglect the horizontal derivatives of the displacement components when compared with the vertical derivatives, and equation (\ref{perpendicular_1}) becomes,

\begin{eqnarray}
\frac{\mu_0\rho_0}{\left|\bm{B}_0\right|^2}\frac{\partial^2 {\bar{\xi}}_{\bot}}{\partial t^2}=\left(1+\frac{B^2_x}{\left|\bm{B}_0\right|^2}\tilde{\beta}\right)\frac{\partial^2 {\bar{\xi}}_{\bot}}{\partial z^2}+\frac{B_x B_z}{\left|\bm{B}_0\right|^2}\tilde{\beta}\frac{\partial^2 {\bar{\xi}}_{||}}{\partial z^2} + \nonumber
%\label{eq:perpendicular_disp}
\end{eqnarray}

\begin{eqnarray}
\ \ \ \ \ \ \ \ \ \ \ \ \  \ \ \ \ \ \ \ \frac{B_x B_z}{\left|\bm{B}_0\right|^2}\frac{\tilde{\beta}}{p_0}\frac{dp_0}{dz}\left(\frac{\partial {\bar{\xi}}_{||}}{\partial z}+\frac{B_x}{B_z}\frac{\partial {\bar{\xi}}_{\bot}}{\partial z}\right),
\label{eq:perpendicular_disp}
\end{eqnarray}
where $\tilde{\beta}=\gamma p_0 \mu_0/|\bm{B}_0|^2$. 

We can combine equations (\ref{eq:parallel_disp}) and (\ref{eq:perpendicular_disp}) and rewrite the system as follows,

\begin{eqnarray}
\frac{\partial^2 {\bar{\xi}}_{||}}{\partial z^2}+\frac{B_x}{B_z}\frac{\partial ^2 {\bar{\xi}}_{\bot}}{\partial z^2}+\frac{d p_0}{p_0 dz}\frac{\partial{\bar{\xi}}_{||}}{\partial z} + \nonumber
\end{eqnarray} 

\begin{equation}
\label{eq: 2.18}
\ \ \ \ \ \ \ \ \ \frac{d p_0}{p_0 dz}\frac{B_x}{B_z}\frac{\partial{\bar{\xi}}_{\bot}}{\partial z}-\frac{\left|\bm{B}_0\right|^2 \rho_0}{B_z^2\gamma p_0}\frac{\partial ^2 {\bar{\xi}_{||}}}{\partial t^2} = 0,
\end{equation} 

\vspace{0.3cm}

\begin{equation}
\label{eq: 2.19}
\frac{\partial ^2 {\bar{\xi}}_{\bot}}{\partial z^2}- \tilde{\beta}\frac{\rho_0}{\gamma p_0}\frac{\partial^2{\bar{\xi}}_{\bot}}{\partial t^2}+\tilde{\beta}\frac{\rho_0}{\gamma p_0}\frac{B_x}{B_z}\frac{\partial^2 {\bar{\xi}}_{||}}{\partial t^2}=0.
\end{equation}

\vspace{0.3cm}

So, together, equations (\ref{eq: 2.18}) and (\ref{eq: 2.19}) govern the displacement components parallel and perpendicular to the direction of the magnetic field, derived in the plane-parallel local analysis and under the approximations presented in this section. 

\subsection{Magnetically dominated region}

There is a close relation between the location of the coupling region in a roAp star and the strength of the magnetic field in that star. The photosphere of a typical roAp star is located in different parts of the surface boundary layer, depending on the magnetic field intensity. For magnetic fields of the order of 1 kG, it is located in the coupling region, while for magnetic fields of the order of 3 kG it is located in the  magnetically dominated region \citep{2007CoAst.150...48C}. Therefore, the oscillations in the atmospheres of these stars might look significantly different from star to star, and in different regions within the same star. In our analysis we are particularly interested in the form of the velocity in the outer atmospheric region. This means that in the region of interest, the magnetic pressure is much larger than the gas pressure and, consequently, the magnetoacoustic wave is decoupled into its acoustic and magnetic components.

Thus, we next consider regions where $\tilde{\beta}$ is much smaller than one, and assume the displacement can be expressed as the sum of a slow (essentially acoustic) component, $\bm{\bar{\xi}}_s$, and a fast (essentially magnetic) component, $\bm{\bar{\xi}}_f$ , i.e.,

\begin{equation}
\label{eq: 2.20}
\bm{\bar{\xi}}=\bm{\bar{\xi}}_s+\bm{\bar{\xi}}_f,  
\end{equation}
such that ${(\bar{\xi}}_{si})^{-1} d{\bar{\xi}}_{si}/dz >> ({\bar{\xi}}_{fi})^{-1} d{\bar{\xi}}_{fi}/dz$, where ${\bar{\xi}}_{si}$ and ${\bar{\xi}}_{fi}$ stand, respectively, for any of the components of the vectors $\bm{\bar{\xi}}_{s}$ and $\bm{\bar{\xi}}_{f}$.
If the two components are indeed decoupled, equations (\ref{eq: 2.18}) - (\ref{eq: 2.19}) must be satisfied separately by $\bm{\bar{\xi}}_s$ and $\bm{\bar{\xi}}_f$. Moreover, we will consider solutions of the form,
\begin{eqnarray}
\bm{\bar{\xi}}(x,z,t)=\bm{\xi}(x,z)e^{(i\omega t)}\nonumber,
\end{eqnarray}
where $\omega$ is the oscillation frequency, so that $\frac{\partial ^2 \bm{\bar{\xi}}}{\partial t^2}=-\omega^2\bm{\bar{\xi}}$.
Finally, we note that, as discussed in the introduction, spectral lines of the rare earth elements form very high in the atmosphere where the it becomes close to isothermal. Since our ultimate aim is to compare the velocity field derived from this analysis with that deduced observationally from the lines of the rare earth elements, we will make an additional approximation and consider only those outermost layers where we may expect the isothermal approximation to be appropriate. Also, we take the first adiabatic exponent, $\gamma$, to be constant in that region, and equal to 5/3.\

In these layers we can define the pressure, $p_0$, and the density, $\rho_0$, as \citep[e.g.][]{1993afd..conf..399G}

\begin{equation}
p_0=\breve{p}_0 e^{-z/H},
\label{pressure}
\end{equation}

\begin{equation}
\rho_0=\breve{\rho}_0 e^{-z/H},
\label{density}
\end{equation}
where $\breve{p}_0$ and $\breve{\rho}_0$ are the pressure and density, respectively, at the base of the isothermal region, and $H$ is the scale height defined by $H^{-1}=-1/p_0dp_0/dz=-1/\rho_0d\rho_0/dz$. 
With these expressions we simultaneously define $z=0$ as being the position of the base of the isothermal region.

\subsubsection{Slow Component}

In the limit when the fast and slow components are decoupled, the approximate form of the space-dependent part of the slow component, $\bm{\xi}_s= {\xi}_{||s}\bm{\hat{e}}_{||}+{\xi}_{\bot s}\bm{\hat{e}}_{\bot}$ can be obtained from the system of equations,
\begin{equation}
\label{eq: 2.26}
\frac{d^2 {\xi}_{||s}}{dz^2}+\frac{dp_0}{p_0dz}\frac{d {\xi}_{||s}}{dz}+\frac{\omega^2 \rho_0}{\gamma p_0}\frac{\left|\bm{B}_0\right|^2}{B_z^2}{{\xi}}_{||s} = 0.
\end{equation}
 \begin{equation}
\label{eq: 2.25}
\frac{d^2 {\xi}_{\bot s}}{d z^2}-\tilde{\beta}\frac{\omega^2 \rho_0}{\gamma p_0}\frac{B_x}{B_z}{\xi}_{||s}=0.
\end{equation}
These are derived from the system of equations (\ref{eq: 2.18}) - (\ref{eq: 2.19}), after neglecting small terms (as discussed in Appendix A).
Moreover, the latter implies that ${\xi}_{\bot s}<<{\xi}_{|| s}$, as expected for a wave that is essentially acoustic in nature, in a magnetically dominated region.

In the present case, of an isothermal atmosphere, equation~(\ref{eq: 2.26}) admits solutions of the type,
\begin{equation}
\label{eq: 2.30c}
\xi_{||s}=p_0^{-1/2}(\tilde{A}e^{ik_{||}z}+\tilde{B}e^{-ik_{||}z}).
\end{equation}  
where,
\begin{equation}
\label{eq: 2.29}
k^2_{||}=\left[\left(\frac{1}{2}\frac{dp_0}{p_0dz}\right)^2-\frac{1}{2p_0}\frac{d^2 p_0}{dz^2}+\frac{\omega^2 \rho_0\left|\bm{B}_0\right|^2}{\gamma p_0 B_z^2}\right],
\end{equation}
and $\tilde{A}$ and $\tilde{B}$ are complex, depth independent, amplitudes.

From equation (\ref{eq: 2.29}) we can derive the acoustic critical frequency, $\omega_{ac}$, defined as the frequency at which $k^2_{||}=0$.  We find, 
\begin{equation}
\label{eq:freq_critica}
\omega_{ac}=\frac{c}{2H}\cos(\alpha_z),
\end{equation}
where $\alpha_z$ is the angle between the direction of the local magnetic field and the local vertical coordinate. If we consider the case of the magnetic pole ($\alpha_z=0$), we find that the latter equation is similar to the cutoff frequency for acoustic waves in the absence of the magnetic field.  Therefore, hereafter we shall use acoustic cutoff frequency for the acoustic critical frequency at the magnetic pole, and retain the name acoustic critical frequency for the critical frequency at a general latitude.

Going back to the solution expressed by equation (\ref{eq: 2.30c}), we  find that when the oscillation frequency is larger than the acoustic critical frequency, $\tilde{A}$ must be set to zero, as otherwise energy would be sent in from outside the star. In that case, we find, from equations~(\ref{eq: 2.25}) and (\ref{eq: 2.30c}), 
\begin{eqnarray}
\label{eq: 2.31a}
{\xi}_{||s}=\frac{\tilde{B}}{p_0^{1/2}}e^{-ik_{||}z} \hspace{0.1cm} ; & & \hspace{-0.5cm}\xi_{\bot s}=\tilde{\beta}\frac{\omega^2 \rho_0}{\gamma p_0^{3/2}}\frac{B_x}{B_z}\frac{\tilde{B}e^{-ik_{||}z}}{(ik_{||}+1/{(2H)})^2}.
\end{eqnarray} 
If we consider the case of an oscillation frequency smaller than the acoustic critical frequency, we find that $\tilde{B}$ must be zero, so that the energy does not increase with $z$. In that case, equations (\ref{eq: 2.25}) and (\ref{eq: 2.30c}) imply that,   
\begin{eqnarray}
\label{eq: 2.31b}
{\xi}_{||s}=\frac{\tilde{A}}{p_0^{1/2}}e^{ik_{||}z} \hspace{0.1cm} ; & & \hspace{-0.5cm}\xi_{\bot s}=\tilde{\beta}\frac{\omega^2 \rho_0}{\gamma p_0^{3/2}}\frac{B_x}{B_z}\frac{\tilde{A}e^{ik_{||}z}}{(-ik_{||}+1/{(2H)})^2}.
\end{eqnarray} 
In both cases, we have set the integration constants that result from the integration of equation~(\ref{eq: 2.25}) to zero, as non-zero constants would correspond to solutions that do not obey the condition set for the slow wave, namely that ${(\bar{\xi}}_{si})^{-1} d{\bar{\xi}}_{si}/dz >> ({\bar{\xi}}_{fi})^{-1} d{\bar{\xi}}_{fi}/dz$.

From equation (\ref{eq:freq_critica}) we can see that $\omega_{ac}$ depends on colatitude, because it depends on the inclination of the magnetic field. So, even when the oscillation frequency is lower than the acoustic cutoff frequency, there will be a colatitude at which, due to the inclination of the magnetic field, the acoustic wave will start to propagate. For this colatitude, above which the acoustic component starts to propagate as a running wave, we define a critical angle, $\theta_{ac}$. 

\subsubsection{Fast Component}

Similarly, in the limit when the fast and slow components are decoupled, the approximate form of the space-dependent part of the fast component, $\bm{\xi}_f= {\xi}_{||f}\bm{\hat{e}}_{||}+{\xi}_{\bot f}\bm{\hat{e}}_{\bot}$ can be obtained from the system of equations,
\begin{equation}
\label{eq: 2.35}
\frac{d^2 \xi_{\bot f}}{d z^2}+\frac{dp_0}{p_0 dz}\frac{d \xi_{\bot f}}{dz}+\frac{\left|\bm{B}_0\right|^2 \omega^2 \rho_0}{B_x B_z \gamma p_0}\xi_{|| f}=0.
\end{equation} 
\begin{equation}
\label{eq: 2.37}
\frac{d^2 \xi_{\bot f}}{d z^2}+k^2_{\bot}\xi_{\bot f}=0,
\end{equation}
where,
\begin{equation}
\label{eq: 1}
k_{\bot}^2=\tilde{\beta}\frac{\omega^2 \rho_0}{\gamma p_0}=\frac{\omega^2}{v^2_{A}},
\end{equation}
and $v_{A}$ is the Alfv\'en velocity.

Equations~(\ref{eq: 2.35}) and (\ref{eq: 2.37}) are derived from the system of equations (\ref{eq: 2.18})-(\ref{eq: 2.19}), by neglecting small terms (see Appendix A for details). An additional conclusion of that analysis is that  ${\xi}_{|| f}<<{\xi}_{\bot f}$, as expected for a wave that is essentially magnetic in nature, in a magnetically dominated region.  Moreover, we note that $k_{\bot}^2$ is always positive, which means that unlike the case of the slow component, in this case there is no magnetic critical frequency. 

%We note that this does not contradict the result found in the paper by \citet{2008MNRAS.tmp..337S}, where the authors present an expression for a magnetic critical frequency. In their paper a different dependant variable was used to describe the fast wave component. Since the critical frequency is derived from a mathematical property of the wave equation, it depends on the variable used to describe the wave  \cite[e.g.][]{1998A&A...337..487S}. However, the physics remains the same, as it should, since to within the approximations made in the two works, the magnetic solution in the magnetically dominated region is the same. From the energetic point of view, the energy contained in the fast component decreases as density decreases in the atmosphere. That led to different different expressions for the critical frequencies were found, illustrates that the turning points are properties of the equations and that depending on the variable used to describe a given phenomena, different expressions for the turning point may be found

In the present case, of an isothermal atmosphere, equation (\ref{eq: 2.37}) admits solutions of the type,
\begin{equation}
\label{eq: 2.39}
\xi_{\bot f}=C_1 J_0\left(\frac{2H\omega \sqrt{\mu_0}}{\left|\bm{B}_0\right|} \rho_0^{1/2}\right)
+C_2 Y_0\left(\frac{2H\omega \sqrt{\mu_0}}{\left|\bm{B}_0\right|} \rho_0^{1/2}\right),
\end{equation}
where $C_1$ and $C_2$ are depth independent amplitudes. Since the function $Y_0$ diverges when its argument tends to zero, $Y_0$ diverges as $\rho_0\rightarrow 0$ and therefore $C_2$ must be zero to obtain a physically meaningful result. 
Moreover, when $\tilde{\beta}$ is sufficiently small, equation (\ref{eq: 2.35}) can be further approximated by,
\begin{equation}
\label{eq: 2.36}
\frac{1}{p_0}\frac{d p_0}{d z}\frac{d \xi_{\bot f}}{d z}\approx\frac{-1}{\tilde{\beta}}\frac{\mu_0 \omega^2 \rho_0}{B_x B_z}\xi_{||f}.
\end{equation}
Thus, we find that the components of the fast displacement are, within the approximations considered,
\begin{eqnarray}
\label{eq: 2.41}
\xi_{\bot f} & = & C_1 J_0\left(\frac{2H\omega \sqrt{\mu_0}}{\left|\bm{B}_0\right|} \rho_0^{1/2}\right)  \nonumber\\ \xi_{|| f} & = & C_1 \frac{\tilde{\beta} B_x B_z}{H \left|\bm{B}_0\right|\omega\rho_0^{1/2}\sqrt{\mu_0}}J_1\left(\frac{2H \omega \sqrt{\mu_0}}{\left|\bm{B}_0\right|}\rho_0^{1/2}\right).
\end{eqnarray}

Since the density $\rho_0$ tends to zero, as one considers higher layers in the atmosphere, $J_0$ tends to one there. Consequently, the solution for the fast component of the displacement perpendicular to the direction of the magnetic field, will tend to a constant.\
Moreover, both $\tilde{\beta}$ and $J_1$ tend to zero, as one considers higher layers in the atmosphere. Thus, the solution for the fast component of the displacement parallel to the direction of the magnetic field will tend to zero, as it should, since the pure magnetic wave will have a displacement perpendicular to $\bm{B}_0$. 
 
\subsection{Dimensionless Equations}

Prior to computing the expected radial velocity, we convert the quantities derived to this point that will enter such computation to corresponding dimensionless quantities. To that end we define:

\vspace{0.25cm}

\ $\eta=z/R$,	\ \ $\sigma=\omega/\omega_0$,	\ \ $p=p_0/\breve{p}_0$, \ \ $\rho=\rho_0/\breve{\rho}_0$,

\vspace{0.25cm}
\ $\varepsilon_{||}=\xi_{||}/R$, \ \ $\varepsilon_{\bot}=\xi_{\bot}/R$,\ \ $C_s= \breve{p}_0/(\breve{\rho}_0 \omega_0^2 R^2)$,.

\vspace{0.25cm}
\ $b_i=B_i/(\mu_0\breve{\rho}_0\omega^2_0 R^2)^{1/2}$,\ \ $b_0=|\bm{B_0}|/(\mu_0\breve{\rho}_0\omega^2_0 R^2)^{1/2}$,
\vspace{0.25cm}
where $R$ is the radius of the star and $\omega_0$=$\sqrt{GM/R^3}$, $M$ is the mass of the star and $G$ is the gravitational constant.
We note that $p$ and $\rho$ defined above are not to be confused with the dimensional quantities named by the same symbol and used in equations (\ref{eq: 2.10})-(\ref{eq: 2.13}). 
With these definitions the dimensionless scale height becomes $\mathcal{H}^{-1}=R/H=-p^{-1}dp/d\eta$. 
Taking these new variables and equations (\ref{eq: 2.29}) and (\ref{eq: 1}), we find the dimensionless $\breve{k}_{||}$ and $\breve{k}_{\bot}$ as,
\begin{equation}
\label{2.42}
\breve{k}_{||}\equiv k_{||}R=\left[\frac{\sigma^2 \rho}{\gamma p C_s \cos^2(\alpha_z)}-\frac{1}{4\mathcal{H}^2}\right]^{1/2},
\end{equation}
\begin{equation}
\label{2.43}
\breve{k}_{\bot}\equiv k_{\bot}R=\left[\frac{\sigma^2 \rho}{b_0^2}\right]^{1/2},
\end{equation}
and we define also the dimensionless acoustic critical frequency, $\sigma_{ac}$, as
\begin{equation}
\sigma_{ac}=\frac{c_d}{2\mathcal{H}}\cos(\alpha_z),
\label{eq:sigma_critico_ac}
\end{equation}
where $c_d$ is the dimensionless sound speed defined as $c_d=(C_s\gamma p/\rho)^{1/2}$. 

With these new variables we derive expressions for the dimensionless slow and fast wave solutions and combine them to obtain the expressions for the dimensionless parallel, $\bar{\varepsilon}_{||}={\varepsilon}_{||} e^{\sigma t}$,  and perpendicular,  $\bar{\varepsilon}_{\bot}={\varepsilon}_{\bot} e^{\sigma t}$, components of the displacement. For oscillations with frequencies larger than $\sigma_{ac}$, we get
\begin{equation}
\label{eq: 2.48a}
\bar{\varepsilon}_{||}=\frac{A_{s}}{p^{1/2}} e^{i(-\breve{k}_{||}\eta+\sigma t)}+\frac{\tilde{\beta}A_{f}b_x b_z}{\mathcal{H}b_0 \sigma \rho^{1/2}}J_1(2\sqrt{\chi\rho})e^{i\sigma t},
\end{equation} 
\begin{eqnarray}
\bar{\varepsilon}_{\bot}= A_{f}J_0(2\sqrt{\chi\rho})e^{i\sigma t}\ + \nonumber
\end{eqnarray}
\begin{equation}
\label{eq: 2.49a}
\ \ \ \ \ \ \ \frac{\tilde{\beta}A_{s}\sigma\rho}{\gamma p^{3/2} C_s}\frac{b_x}{b_z}\left[\frac{2 \mathcal{H}}{1+2i\mathcal{H}\breve{k}_{||}}\right]^2 e^{i(-\breve{k}_{||}\eta+\sigma t)},
\end{equation}
and for oscillations with frequencies smaller than $\sigma_{ac}$, we get
\begin{equation}
\label{eq: 2.48b}
\bar{\varepsilon}_{||}=\frac{A_{s}}{p^{1/2}} e^{i(\breve{k}_{||}\eta+\sigma t)}+\frac{\tilde{\beta}A_{f}b_x b_z}{\mathcal{H}b_0 \sigma \rho^{1/2}}J_1(2\sqrt{\chi\rho})e^{i\sigma t},
\end{equation} 
\begin{eqnarray}
\bar{\varepsilon}_{\bot}= A_{f}J_0(2\sqrt{\chi\rho})e^{i\sigma t} \ +\nonumber
\end{eqnarray}
\begin{equation}
\label{eq: 2.49b}
\ \ \ \ \ \ \ \frac{\tilde{\beta}A_{s}\sigma\rho}{\gamma p^{3/2} C_s}\frac{b_x}{b_z}\left[\frac{2 \mathcal{H}}{1-2i\mathcal{H}\breve{k}_{||}}\right]^2 e^{i(\breve{k}_{||}\eta+\sigma t)},
\end{equation}
where $A_s$ is a complex, depth independent amplitude which depends on latitude, defined in terms of the previously adopted amplitudes by $A_s=\tilde{A}/(R\sqrt{\breve{p}_0})$ or $A_s=\tilde{B}/(R\sqrt{\breve{p}_0})$, depending on the colatitude. Similarly, $A_{f}$ is a complex depth independent amplitude which depends on colatitude, $\chi=\mathcal{H}^2\sigma^2/\left|b_0\right|^2$, and $t$ is the dimensionless time. 

The complex amplitudes $A_s$ and $A_f$ can be written in terms of their modulus, $|A_s|$ and $|A_f|$, and phases, $\phi_s$ and $\phi_f$. Before we proceed to calculate the disk integrated line-of-sight projected velocity, we still need to determine these amplitudes and phases. Their relative values are not arbitrary. Instead, they are imposed by the magnetoacoustic coupling that takes place below, in the region where the magnetic and gas pressure are comparable. Thus, in order to find their values we fit our expressions for the displacement to the numerical solutions that come out of the MAPPA code \citep{2006MNRAS.365..153C}. From the fit, we extract the amplitudes and phases and, simultaneously, confirm that our analytical solutions for the displacement represent well the corresponding numerical solutions, in the limit when both  $\tilde{\beta}$ is much smaller than one, and the atmosphere becomes isothermal.

Generally, since in the region of interest, $\tilde{\beta}$ is much smaller than one, we may expect that the second term on the right hand side of equations~(\ref{eq: 2.48a})-(\ref{eq: 2.49b}) can be neglected.
That assumption, however, can break down if the relative value of the amplitudes $|A_s|$ and $|A_f|$ is such as to compensate for the small value of $\tilde{\beta}$. We checked whether that was the case and found that the only situation in which the two terms become of the same order of magnitude is in equation (\ref{eq: 2.48a}), for colatitudes very close to 90$\arcdeg$. In those cases, we have included the second term in our computations. In all other cases the second terms on the right hand side of equations~(\ref{eq: 2.48a})-(\ref{eq: 2.49b}) were neglected.\

\section{\label{RV}Theoretical Radial Velocity}

	In the observations what is seen is an integral over the visible stellar disk. Therefore, we shall integrate the line-of-sight projection of the velocity over the visible stellar disk taking into account the limb darkening effect, to simulate what observers would see if they had the means to measure the radial velocity as function of atmospheric depth.
	
\begin{figure}
\includegraphics[width=20pc]{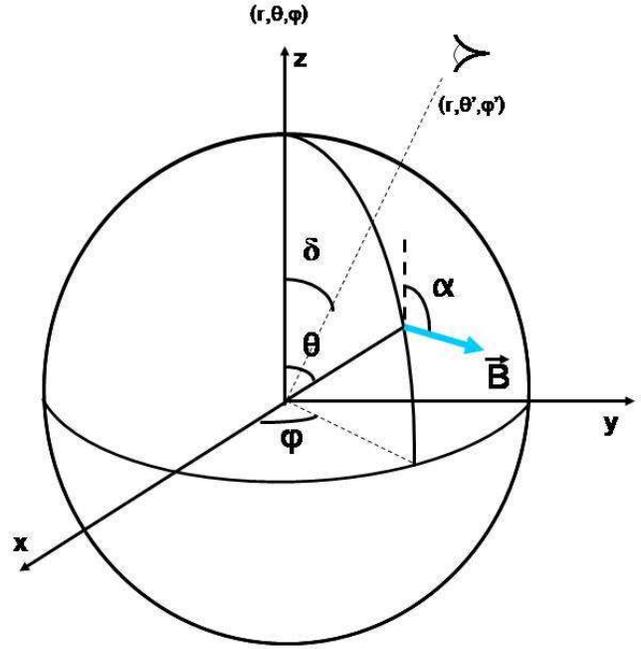}\hspace{4pc}
\caption{Schematic representation of the star. $\bm{B}$ is the local magnetic field vector, $\alpha$ is the angle between the magnetic field axis and the local magnetic field vector and $\delta$ the angle between the magnetic field axis and the direction of the observer.}
\label{fig.estrela_geral}
\end{figure}

	In Fig.\ref{fig.estrela_geral} we show a schematic representation of the star, where the observer is considered to be at a general position. In the figure $\delta$ is the angle between the magnetic field axis and the direction of the observer, which we shall assume to be fixed, meaning that we are considering the observer's view at a particular rotation phase. Moreover, $\bm{B}$ is the local magnetic field vector and $\alpha$ is the angle between the magnetic field axis and the direction of the local magnetic field. In our analysis we consider two spherical coordinate systems: The spherical coordinate system, (r, $\theta$, $\varphi$), aligned with the magnetic field axis, and the spherical coordinate system aligned with the direction of the observer, (r, $\theta'$,$\varphi'$). The coordinates in the two systems are related through standard expressions given, e.g., by \citet{1996fuas.book.....K}.
  
 Taking the real part of equations (\ref{eq: 2.48a}) and (\ref{eq: 2.48b}), after neglecting the second terms on the right hand side, and differentiating with respect to time, we can write a general expression for the component of the velocity parallel to the direction of the magnetic field in the form,
\begin {eqnarray}
v_{||}(\eta,\theta,t)=-\sigma\frac{A_{s}}{p^{1/2}}e^{(-q_e |\breve{k}_{||}|\eta) }\sin(-q_r |\breve{k}_{||}|\eta+\sigma t)\nonumber,
\end{eqnarray}
where $q_e$ and $q_r$ are factors defined in the following way: when the oscillation frequency is larger than the acoustic critical frequency, $q_e=0$ and $q_r=1$; when the oscillation frequency is smaller than the acoustic critical frequency, $q_e=1$ and $q_r=0$.
	Similarly, from equations (\ref{eq: 2.49a}) and (\ref{eq: 2.49b}), after neglecting the second terms on the right hand side, we find that the component of the velocity perpendicular to the direction of the magnetic field is given by, 
\begin {eqnarray}
v_{\bot}(\eta,\theta,t)=-\sigma A_{f}J_0(2\sqrt{\chi\rho})\sin(\sigma t)\nonumber.
\end{eqnarray}

Assuming a linear limb darkening law, with $a$ as the limb darkening coefficient, the expression for the velocity component parallel to the line of sight averaged over part of the visible stellar disk for a general position of the observer,  $v_{int}$, thus becomes, 
\begin{eqnarray}
v_{int}(\delta,\eta,t)=\int_{\varphi'_i}^{\varphi'_f}\int_{\theta'_i}^{\theta'_f}\big[-\frac{A_{s}}{p^{1/2}}e^{(-q_e |\breve{k}_{||}|\eta) }\times \nonumber 
\end{eqnarray}
\begin{eqnarray}
\sin(-q_r |\breve{k}_{||}|\eta+\sigma t)X_{||}-A_{f}J_0(2\sqrt{\chi\rho})\sin(\sigma t)X_{\bot}\big] \nonumber
\end{eqnarray}
\begin{eqnarray}
\label{eq: 2.55}
\ \ \ \ \ \ \ \ \ \ \ \ \ \ \ \ \ \ \times\frac{\sigma}{2C_n}(1-a(1-\cos\theta'))\sin(2\theta')d\theta' d\varphi',
\end{eqnarray}
\vspace{0.06cm}
where $\varphi'_i$, $\varphi'_f$, $\theta'_i$ and $\theta'_f$ are, respectively, the limit in longitude and latitude, in the observer's spherical coordinate system, of the region of the visible stellar disk over which the integration is to be made. For an integration over the whole visible stellar disk $\theta'$ varies within the interval $[0,\pi/2]$ and $\varphi'$ varies within the interval $[0,2\pi]$. Moreover, $C_n$ is a normalization factor defined as,
\vspace{0.06cm}

\begin{eqnarray}
\label{eq: 2.56}
C_n=\int_{\varphi'_i}^{\varphi'_f}\int_{\theta'_i}^{\theta'_f}(1-a(1-\cos\theta'))\sin(2\theta')d\theta'd\varphi', \nonumber
\end{eqnarray}
\vspace{0.6cm}
and $X_{||}$ and $X_{\bot}$ are the line-of-sight projections of the unit vectors $\bm{\hat{e}}_{||}$ and $\bm{\hat{e}}_{\bot}$, respectively.

\subsection{Fitting to Acos($\sigma$t+$\phi$)}

	 In spectroscopic observations the radial velocity derived from the time series analysis is usually fitted to a function of the form Acos($\sigma$t+$\phi$). From this fit, the authors derive the amplitude and the phase of the oscillations \citep[e.g.][]{2005MNRAS.358L...6K,2006MNRAS.370.1274K,2007A&A...473..907R,2007CoAst.150...81S}.\
	 Since our ultimate goal is to compare the results of our work with the observations, in our study we fit the average line-of-sight velocity, $v_{int}$, to a function similar to the one above, and derive the amplitudes and phases of the oscillations as function of height in the atmosphere. 
From the fit we find the expression for the fitted phase, $\phi$, namely,

\begin{eqnarray}
\phi(\delta,\eta)=\arctan\big(\big[\int_{\varphi'_i}^{\varphi'_f}\int_{\theta'_i}^{\theta'_f}[C_{||}\cos(-q_r |\breve{k}_{||}|\eta+\phi_{s})e^{-q_e |\breve{k}_{||}|\eta}\nonumber
\end{eqnarray}

\begin{eqnarray}
+C_{\bot}\cos(\phi_{f})]d\theta'd\varphi'\big]\times\big[\int_{\varphi'_i}^{\varphi'_f}\int_{\theta'_i}^{\theta'_f}[-C_{||}\sin(-q_r |\breve{k}_{||}|\eta+\phi_{s})\nonumber
\end{eqnarray}

\begin{eqnarray}
\label{eq: 2.59}
e^{-q_e |\breve{k}_{||}|\eta}-C_{\bot}\sin(\phi_{f})]d\theta'd\varphi'\big]^{-1}\big),
\end{eqnarray}
\vspace{0.3cm}
and the expression for the fitted amplitude, A, namely,

\begin{eqnarray}
{\rm A(\delta,\eta)}=\frac{\sigma}{\sqrt{2}C_n}\big[\big[\int_{\varphi'_i}^{\varphi'_f}\int_{\theta'_i}^{\theta'_f}
[-C_{||}\sin(-q_r |\breve{k}_{||}|\eta+\phi_{s})\nonumber
\end{eqnarray}

\begin{eqnarray}
e^{-q_e |\breve{k}_{||}|\eta} -C_{\bot}\sin(\phi_{f})]d\theta'd\varphi'\big]^2+\big[\int_{\theta'_i}^{\theta'_f}\int_{\varphi'_i}^{\varphi'_f}[C_{||}e^{-q_e |\breve{k}_{||}|\eta}\nonumber
\end{eqnarray}

\begin{eqnarray}
\label{eq: 2.60}
\cos(-q_r |\breve{k}_{||}|\eta+\phi_{s}) +C_{\bot}\cos(\phi_{f})]d\theta'd\varphi'\big]^2\big]^{1/2},
\end{eqnarray}
\vspace{0.3cm}
where $C_{||}$ and $C_{\bot}$ are given by,

\vspace{0.15cm}
\begin{eqnarray}
\label{eq: 2.61}
C_{||}=\frac{|A_{s}|}{2p^{1/2}}(1-a(1-\cos\theta'))\sin(2\theta')X_{||},
\end{eqnarray}
and
\begin{eqnarray}
\label{eq: 2.62}
C_{\bot}=\left|A_{f}\right|J_0(2\sqrt{\chi}\rho)(1-a(1-\cos\theta'))\sin(2\theta')X_{\bot}.
\end{eqnarray}

\subsection{A toy Model}
\label{toy model}
	In sec.~\ref{results} we will use expressions (\ref{eq: 2.59}) to (\ref{eq: 2.62}) to study the behavior of the phase and amplitude of $v_{int}$ in a number of case studies. Here we try to anticipate those results through an analytical analysis of the sum of wave-like functions.  This simple analysis will help support the interpretation of the results obtained in the case studies. 

The motivation for the form of the wave-like functions that we shall consider comes from equation (\ref{eq: 2.55}). In the case of an observer aligned with the magnetic field axis, this equation can be written as the sum of three integrals, namely,
\begin{eqnarray}
\label{v_int_sep}	
v_{int}(\delta,\eta,t)=I_1+I_2+I_3, 
\end{eqnarray}
where,
\begin{eqnarray}	
I_1\tiny{=-\int_{\varphi'_i}^{\varphi'_f}\int_{\theta'_i}^{\theta'_c}\frac{\sigma}{C_n}e^{-|\breve{k}_{||}|\eta}\sin(\sigma t\tiny{+}\phi_s) C_{||} d\theta' d\varphi'},
\label{evanescente}
\end{eqnarray}
\begin{eqnarray}	
I_2\tiny{=-}\tiny{\int_{\varphi'_i}^{\varphi'_f}\int_{\theta'_c}^{\theta'_f}}\frac{\sigma}{C_n}\sin(\tiny{-|\breve{k}_{||}|}\eta\tiny{+}\sigma t\tiny{+} \phi_{s}) C_{||} d\theta'd\varphi',
\label{running}
\end{eqnarray}
\begin{eqnarray}	
I_3\tiny{=-\int_{\varphi'_i}^{\varphi'_f}\int_{\theta'_i}^{\theta'_f}}\frac{\sigma}{2C_n}\sin(\sigma t\tiny{+}\phi_{f}) C_{\bot} d\theta' d\varphi',
\label{standing}
\end{eqnarray}
where $\theta'_c = \theta_{ac}$ is the critical angle above which the acoustic component will start to propagate as a running wave. Thus, $I_1$ contains the information of the evanescent acoustic solution only, $I_2$ contains the information of the running acoustic solution only, and $I_3$ contains the information of the standing magnetic solution only. In the above we have assumed  $\theta'_i \le\theta'_c\le\theta'_f$. When  $\theta'_c <\theta'_i$, $I_1$ is zero and the lower colatitude limit in $I_2$ is taken to be  $\theta'_i$. Like wise,  when  $\theta'_c >\theta'_f$, $I_2$ is zero and the upper colatitude limit in $I_1$ is taken to be  $\theta'_f$. \
 
\subsubsection{\label{different nature}Visual superposition of waves of different nature}
	
In this toy study we define three simple waves, with velocities $v_1$, $v_2$ and $v_3$, inspired, respectively, by $I_1$, $I_2$, and $I_3$. The velocities are assumed to have the following form, with the subscripts $r$ and $i$ indicating the real and imaginary parts of the wavenumber, respectively,
\begin{eqnarray}	
v_j=\tilde{\psi}_j\sin(-k_{j_r}\eta+\sigma t + \phi_j),
\label{v_1}
\end{eqnarray}
with $j\in \left\{1,2,3\right\}$, and amplitudes and wavenumbers characterized as follows:	
\begin{eqnarray}	
\ \ \ \ \ \tilde{\psi}_1=\psi_1 e^{\eta/2\mathcal{H}}e^{-k_{1_i}\eta} \ \ \ ; \ k_{1_i}=k_1 \ \ \ ;  \ \ \ k_{1_r}=0,
\label{psi_1}
\end{eqnarray}
\begin{eqnarray}	
\ \ \ \ \ \tilde{\psi}_2=\psi_2 e^{\eta/2\mathcal{H}}e^{-k_{2_i}\eta} \ \ \ ; \ k_{2_i}=0 \ \ \ ; \ k_{2_r}=k_2 ,
\label{psi_2}
\end{eqnarray}
\begin{eqnarray}	
\ \ \ \ \ \tilde{\psi}_3= \psi_3e^{-k_{3_i}\eta}\ \ ;  \ \ \ k_{3_i} \ = \ k_{3_r}=0,
\label{psi_3}
\end{eqnarray}
where $\psi_1$, $\psi_2$, $\psi_3$, $k_1$, $k_2$ are all real, constant, and positive. 	

Equations (\ref{v_1}) and (\ref{psi_1}) define a wave-like function that retains the characteristics of the integral of the line-of-sight acoustic component of the velocity when the oscillation frequency is smaller than the acoustic critical frequency, equations (\ref{v_1}) and (\ref{psi_2}) define a wave-like function that retains the characteristics of the integral of the line-of-sight acoustic component of the velocity when the oscillation frequency is larger than the acoustic critical frequency, and equations (\ref{v_1}) and (\ref{psi_3}) define a wave-like function that retains the characteristics of the integral of the line-of-sight magnetic component of the velocity. \
	
	With these three simple wave-like functions in hand, we start by considering the simple case in which one of the contributions can be neglected when compared to the other two. In that case we define our total velocity as,
\begin{eqnarray}	
v_{toy}= v_j + v_k   
\label{vtoy}
\end{eqnarray}
where $j$ and $k$ $\in \left\{1,2,3\right\}$ are indexes that depend on the type of contribution that we are considering in the sum.  

Similar to what we did to $v_{int}$ and to what is done in high-resolution spectroscopic observations, we fitted the equation for the velocity, $v_{toy}$, to an expression of the type $A_{toy}\cos(\sigma t+\phi_{toy})$, and from the fit we derive the expression for the phase, 
\begin{eqnarray}
\phi_{toy}=\arctan\big[(\tilde{\psi}_j\cos(-k_{j_r}\eta+\phi_{j})+\tilde{\psi}_k\cos(-k_{k_r}\eta+\phi_k))\nonumber \\
(-\tilde{\psi}_j \sin(-k_{j_r} \eta+ \phi_j)-\tilde{\psi}_k \sin(-k_{k_r} \eta+\phi_k))^{-1} \big],
\label{eq: 2.1.7}
\end{eqnarray}	          
and for the amplitude,
\begin{equation}
\label{eq: 2.1.6}
A_{toy}^2=(\tilde{\psi}_j-\tilde{\psi}_k)^2+4\tilde{\psi}_j\tilde{\psi}_{k}\cos^2\left[\frac{-(k_{j_r}-k_{k_r})\eta+\phi_{j}-\phi_{k}}{2}\right].
\end{equation}	

These expressions are analogous to equations (\ref{eq: 2.59}) and (\ref{eq: 2.60}), respectively, except that they are substantially simpler because they do not contain the latitudinal dependence of each of the integral components.
 
 Concerning the amplitude, $A_{toy}(\eta)$, we can see it will be affected by the relative size of the amplitudes  $\tilde{\psi}_j$ and $\tilde{\psi}_k$ and by the phase $\left[\frac{-(k_{j_r}-k_{k_r})\eta+\phi_{j}-\phi_{k}}{2}\right]$. If $\tilde{\psi}_j$ is much larger, or much smaller, than $\tilde{\psi}_k$, the first term on the right hand side of equation (\ref{eq: 2.1.6}) will dominate and no structure in the amplitude as function of atmospheric height is expected. Thus the behaviour of the latter with atmospheric height will be similar to the behaviour of the predominant component. That is the most common situation. On the other hand, if the two components have similar amplitudes, i.e., $\tilde{\psi}_j \approx \tilde{\psi}_k$, then the second term on the right hand side of equation (\ref{eq: 2.1.6}) may dominate and some structure in the amplitude as function of atmospheric height may be expected. However that structure will depend strongly on the phase, $\left[\frac{-(k_{j_r}-k_{k_r})\eta+\phi_{j}-\phi_{k}}{2}\right]$, in the region under consideration. In fact, when values typical of those found in our general analysis are taken for the latter, we find that in most cases no structure at all is found, i.e., the amplitude simply increases with atmospheric height as in the first situation considered.     

 The results for the phase, $\phi_{toy}$, will also depend on the relative amplitude of each component and on the argument of the trigonometric functions. However, in this case no matter how large the value for the amplitude of one of the components is, if its phase is such that either the cosine or the sine of the corresponding argument goes through a zero, the component with smaller amplitude will be the one dominating the corresponding sum (nominator or denominator) in equation~(\ref{eq: 2.1.7}). This, as we will see later, will result in a diversity of phase behaviour that will be possible even when one component is much larger than the others.

 For completeness, and because this analytical analysis will be of relevance for the interpretation of the results presented in section \ref{results}, below we provide the expressions for the phase and amplitude obtained in the case of adding three simple wave-like functions, $v_{toy}=v_j+v_k+v_p$. In that case we find,
\begin{eqnarray}
\phi_{toy}=\arctan \big[\big(\tilde{\psi}_{j}\cos(-k_{j_r}\eta+\phi_{j})+\tilde{\psi}_{k}\cos(-k_{k_r}\eta+\phi_{k})\nonumber \\
+\tilde{\psi}_{p}\cos(-k_{p_r}\eta+\phi_{p})\big)\big(-\tilde{\psi}_{j} \sin(-k_{j_r}\eta+\phi_{j})\nonumber \\
-\tilde{\psi}_{k}\sin(-k_{k_r}\eta+\phi_{k})+\tilde{\psi}_{p}\sin(-k_{p_r}\eta+\phi_{p})\big)^{-1}\big],
\label{eq: phi_3}
\end{eqnarray}	          
and 
 \begin{eqnarray}
A_{toy}^2=\big(\tilde{\psi}_{j}\sin(-k_{j_r}\eta+\phi_{j})+\tilde{\psi}_{k}\sin(-k_{k_r}\eta+\phi_{k})+\nonumber \\
\tilde{\psi}_{p}\sin(-k_{p_r}\eta+\phi_{p})\big)^2+\big(\tilde{\psi}_{j}\cos(-k_{j_r}\eta+\phi_{j})+\nonumber \\
\tilde{\psi}_{k}\cos(-k_{k_r}\eta+\phi_{k})+\tilde{\psi}_{p}\cos(-k_{p_r}\eta+\phi_{p})\big)^2.
\label{eq: amp_3}
\end{eqnarray}	

\subsubsection{\label{similar nature}Visual superposition of waves of similar nature}
 Based on equation (\ref{v_int_sep}) we have suggested three components for the velocity, inspired, in particular, in the two solutions for the acoustic component of the velocity and in the solution for the magnetic component. However, if we look at the arguments of expressions (\ref{evanescente}), (\ref{running}) and (\ref{standing}), we notice that the amplitudes, the vertical wavenumbers and the phases all depend on colatitude. This was not considered in section~\ref{different nature} because our intention was to eliminate the integrals, to keep the analysis simple. Nevertheless, it is clear that such dependence can itself introduce a variety of phase behaviour, even when a component of a particular nature is considered alone. To explore this fact, we can associate $v_j$ and $v_k$ in equation~(\ref{vtoy}) to two components of similar nature, with different amplitudes, vertical wavenumbers and input phases. 

As an example, let us consider a single evanescent wave. The amplitude of that wave increases with atmospheric height, while the phase stays constant. On the other hand, if we have a superposition of two evanescent waves with different wave numbers and phases, we can find a phase varying with atmospheric height. Similarly, if we consider a single outwardly running wave, the amplitude increases and the phase decreases with height. However, if we have a superposition between outwardly running waves with different wave numbers and phases, we can have a phase scenario different from the latter.  
 	Finally, when considering a single standing solution defined by equations~(\ref{v_1}) and (\ref{psi_3}), both amplitude and phase are constant with atmospheric height. In this case, if we have a superposition of two of these solutions with different amplitudes and phases, the only thing that may be affected is the sign of the phase, since in all cases the total amplitude and phase will remain constant.\

\section{Results and Analysis}
\label{results}

In this section we present a selection of the results obtained in our study of the integrated line-of-sight velocity. The cases to be presented were chosen so as to illustrate different behaviours of the phase. However, as our aim was to conduct an in-depth study of these cases, we have restricted the parameter space considerably. In particular, we have fixed the position of the observer, which is always considered to be aligned with the magnetic axis, and we have considered modes which in the absence of a magnetic field would be pure dipole modes ($\textit{l}=1$).  The results of a systematic study of the parameter space, including variations of $\delta$ and $\textit{l}$, conducted with the same approach, will be presented in a separate paper. A list of the case studies is shown in Table \ref{cases}.

\begin{table}
\begin{center}
\begin{tabular}{ l l l l}
\hline
\ \ \ \ \ \ \ \ \ \ \ \  &    $B_{p}$ (kG)     &      $\nu$ (mHz)     &      Limits of integration\\
\hline
Case 1\ \ \ \ \   &  \ \ \   4            &   \  1.0     &     whole stellar disk\\
Case 2            &  \ \ \   4            &   \  2.5     &     whole stellar disk\\
Case 3            &  \ \ \   1            &   \  1.7     &     whole stellar disk\\
Case 4            &  \ \ \   1            &   \  2.1     &     whole stellar disk\\
Case 5            &  \ \ \   2            &   \  3.0     &     whole stellar disk\\
Case 6            &  \ \ \  4            &    \ 2.2         &   \ \ \ \ [62\arcdeg,85\arcdeg]\\
\hline\\
\end{tabular}
\caption{Characteristics of the Case Studies. All cases are for an observer pole-on and $l=1$.}
\label{cases}
\end{center}
\end{table}

 In the computations we have used a stellar model taken from the Code d'Evolution Stellaire Adaptatif et Modulaire (CESAM code), \citep[e.g.][]{2008Ap&SS.316...61M}, with a mass $M=1.8$~M$_{\odot}$, a radius $R=1.57$~R$_{\odot}$, and an effective temperature $ T_{\rm eff}=8363$~K.  To this model, we have matched an isothermal atmosphere at a temperature $T_{\rm iso}=6822$~K. The acoustic cut-off frequency of the model is 2.7~mHz. We have considered different values for the polar magnetic field, $B_{p}$, different cyclic oscillation frequencies, $\nu$, varying from $1.0$ mHz to $3.0$~mHz, and different integration intervals, to take into account the possibility that the lines from which the radial velocities are derived in the observations are not homogeneously distributed over the stellar surface but, instead, concentrated in particular regions of the star. \
   
   For each case we derived the amplitude, A, and the phase, $\phi$, of the integrated line-of-sight velocity, using equations (\ref{eq: 2.59}) and (\ref{eq: 2.60}), respectively. Moreover, we fitted equation (\ref{evanescente}) to a function of the type A$_{ev}\cos(\sigma t+\phi_{ev})$ to derive the amplitude, A$_{ev}$, and the phase, $\phi_{ev}$ of the evanescent component; we fitted equation (\ref{running}) to a function of the type A$_{r}\cos(\sigma t+\phi_r)$ to derive the amplitude, A$_{r}$, and the phase, $\phi_r$ of the running component; and we fitted equation (\ref{standing}) to a function of the type A$_{st}\cos(\sigma t+\phi_{st})$ to derive the amplitude, A$_{st}$, and the phase, $\phi_{st}$ of the standing component. In addition, we fitted the total acoustic term of $v_{int}$ to a function of the type A$_{ac}\cos(\sigma t+ \phi_{ac})$ to derive the corresponding amplitude, A$_{ac}$, and phase, $\phi_{ac}$. In Table \ref{represen} we show the different representations of the amplitude and phase according to the component of 
the velocity integral from which they were derived. 

\begin{table}
\begin{center}
\begin{tabular}{ l l l l}
\hline
\ \ \ \ \ \ \ \ \ \ \ \  &   Amplitude     &      Phase     &      Nature\\
\hline
Total velocity\ \ \ \ \  &       A            &     $\phi$     &     Ac + Mag\\
Evanescent component     &       A$_{ev}$     &     $\phi_{ev}$     &     Ac\\
Running component        &       A$_{r}$      &     $\phi_r$     &     Ac\\
Standing component       &       A$_{st}$     &     $\phi_{st}$     &     Mag\\
Total acoustic component &       A$_{ac}$     &     $\phi_{ac}$     &     Ac\\
\hline\\
\end{tabular}
\caption{Different representations for the amplitude (second column), phase (third column), and nature of the wave (where Ac stands for acoustic and Mag stands for magnetic), according to the component of the velocity from which they were derived (first column).}
\label{represen}
\end{center}
\end{table}

%\subsection{Case studies}
%\label{Case studies}

In the plots below, the amplitudes and phases are shown as a function of atmospheric height, given in units of the dimensionless scale height. We note that the zero of the atmospheric height corresponds to the base of the isothermal atmosphere.  In practice, the region where rare earth lines are formed in roAp stars will be only a small slice of the layers that we are considering.  However, the phase of each component contributing to $v_{int}$ in that region will depend on the model considered, as well as on the magnetic field intensity. Since exploring a large grid of magnetic models is out of the scope of this work, we extend the isothermal part of our model atmosphere for 10 scale heights, in the hope of finding a variety of combinations of  $\phi_{st}$, $\phi_r$, and $\phi_{ev}$, and, thus, all possible behaviours of the total phase. 

\subsection{Pulsation Amplitude}
\label{amplitudes}

\begin{figure}
\begin{center}
\includegraphics[width=18pc]{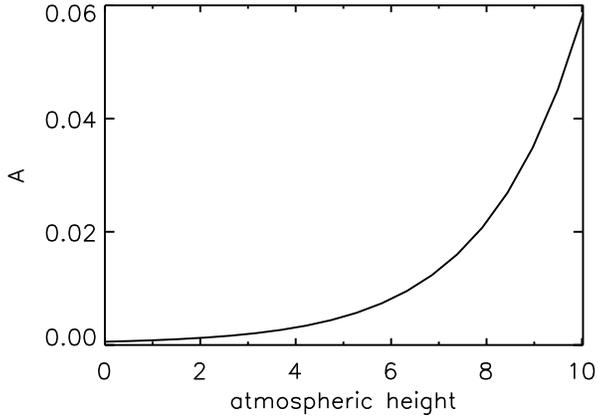}
\caption{Dimensionless amplitude as a function of atmospheric height (in units of $\mathcal{H}$) - $B_{p}$=2 kG,\ $\nu=3.0$ mHz, and for an integration over the visible stellar disk.}
\label{fig:amplitude}
\end{center}
\end{figure}

\begin{figure}
\begin{center}
\includegraphics[width=18pc]{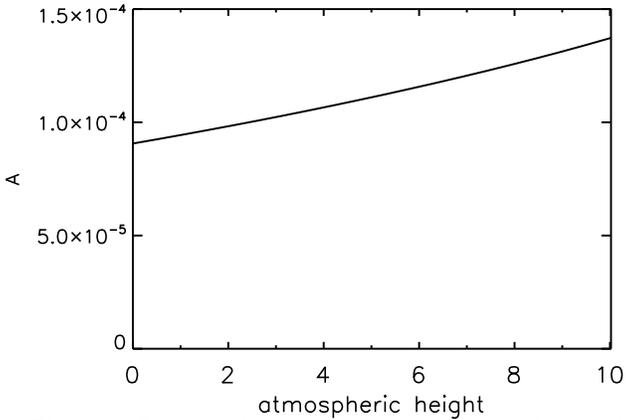}
\caption{Dimensionless amplitude as a function of atmospheric height (in units of $\mathcal{H}$) - $B_{p}$=4 kG,\ $\nu=1.0$ mHz, and for an integration over the visible stellar disk.}
\label{fig:amplitude_eva}
\end{center}
\end{figure}

Concerning the amplitude, we found that in most cases the latter increases with atmospheric height. In Fig.~\ref{fig:amplitude} we show an example of this behavior, for the case of $B_p$=2 kG, $\nu=3.0$ mHz, and an integration over the whole visible stellar disk. This behavior for the amplitude was to be expected given our findings in the toy study presented in section \ref{toy model} and can be explained by the fact that the density is decreasing with height. In this case, the amplitude goes with $1/\sqrt{\rho}$. This is consistent with the running acoustic component dominating the integral of the velocity. 
	In Fig.~\ref{fig:amplitude_eva} we show the amplitude derived for the case of $B_p$=4 kG, $\nu=1.0$ mHz, and an integration over the whole visible stellar disk. While the amplitude in this case also increases with atmospheric height, it does so at a smaller rate than in the previous case. That is a consequence of the integral for the velocity being dominated by the evanescent, rather than the running, acoustic component.

The only exception to this growing behaviour of the amplitude that shall be shown in the present paper is that discussed in the case study 6, presented below. Other exceptions are found when the parameter space is not as restrictive as the one considered here. Those cases shall be discussed in the upcoming paper.

\subsection{Pulsation Phase}
\label{phase}

Concerning the phases, we find a situation that is much more diverse. This was also to be expected given our findings in section \ref{toy model}. 

\vspace{0.5cm}

\underline{Case 1: Constant phase}

\vspace{0.5cm}

	In case 1 we consider a magnetic field with polar magnitude $B_p$=1~kG, and a cyclic oscillation frequency  $\nu=1.0$~mHz. Moreover, the integration is over the whole visible stellar disk.\   
	The oscillation frequency in this case is smaller than the acoustic cutoff frequency. The critical angle, $\theta_{ac}$, is 78\arcdeg, which means that for colatitudes smaller than this angle the acoustic term will be an evanescent wave, with a phase independent of atmospheric height, while for larger colatitudes the acoustic term will be an outwardly running wave with phase decreasing with atmospheric height. 

\begin{figure}
\begin{center}
\includegraphics[width=18pc]{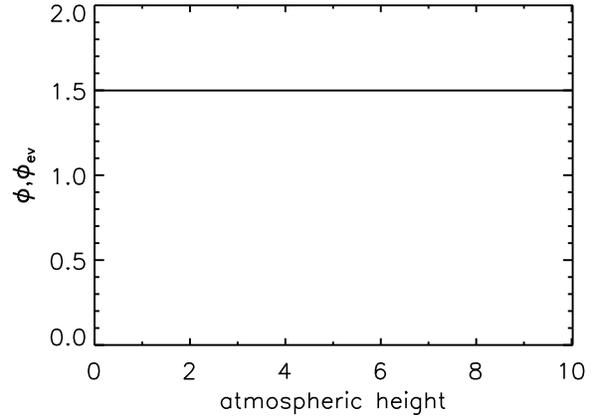}
\caption{Case 1 - $B_{p}$=4 kG,\ $\nu=1.0$ mHz, and integration over the whole visible stellar disk. Total phase overlapped with phase derived from the evanescent term alone as a function of atmospheric height (in units of $\mathcal{H}$). The two lines are coincident.}
\label{fig:phase_caso1}
\end{center}
\end{figure}

\begin{figure}
\begin{center}
\includegraphics[width=20pc]{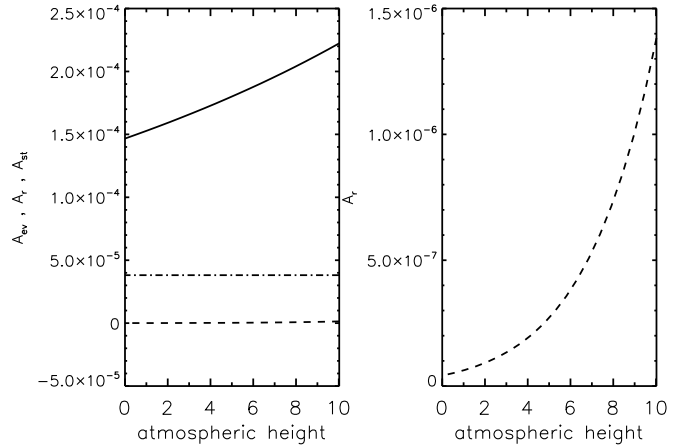}
\caption{Case 1 - same as in figure~\ref{fig:phase_caso1}. 
Left panel: Dimensionless amplitudes derived from each term of the velocity: full line - A$_{ev}$; dashed line - A$_{r}$; dashed-dotted line - A$_{st}$
Right panel: Zoom in the y axis for A$_{r}$.}
\label{fig:amp_sob_caso1}
\end{center}
\end{figure}

As we can see in Fig.\ref{fig:phase_caso1}, the phase obtained with equation (\ref{eq: 2.59}) is constant throughout the region of the atmosphere under consideration. In Fig. \ref{fig:amp_sob_caso1} we show the amplitudes derived from each of the components of the velocity. Clearly, the amplitude derived from the evanescent term alone is always much larger than the amplitudes derived from the running or magnetic terms alone. 

In Fig.\ref{fig:phase_caso1}, we also show the phase derived from the evanescent term alone. It overlaps with the line for the total phase. This is a simple case, the simplest of all cases that we are going to consider here, in which one of the terms of the velocity dominates entirely the final result, due to its large amplitude when compared to the amplitudes exhibited by the other terms. We note, however, that having a dominating amplitude is not a sufficient condition to dominate entirely the final results, as we will see in other examples.\
 
It is important to notice that other situations exist in which the resulting phase is constant due to reasons different to the one considered in this case study. If evanescent or standing waves dominate the integral of the velocity, and if the phases derived from the corresponding terms alone are constant, the resulting phase will generally be constant. Moreover, the resulting phase can sometimes be constant if the visual superposition happens between evanescent and standing waves with similar amplitudes.   

\vspace{0.5cm}

\underline{Case 2: Phase increasing deeper and decreasing higher} 
\underline{in the atmosphere}

\vspace{0.5cm}

The case that we present here is for $B_p$=4~kG, $\nu=2.5$ mHz, and an integration over the whole visible stellar disk.

\begin{figure}
\begin{center}
\includegraphics[width=18pc]{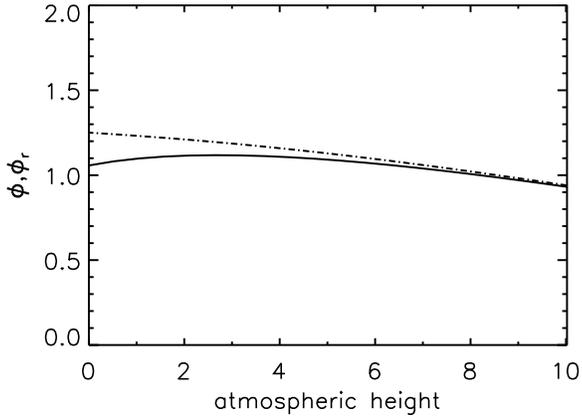}
\caption{Case 2 - $B_{p}$=4 kG,\ $\nu=2.5$ mHz, and an integration over the visible stellar disk. Total phase (full line) and phase derived from the acoustic term alone (dashed-dotted line) as a function of atmospheric height (in units of $\mathcal{H}$).}
\label{fig:phase_caso2}
\end{center}
\end{figure}

	This case is similar to the previous one in the sense that the oscillation frequency is smaller than the acoustic cutoff frequency. However, here the critical angle is only 43$\arcdeg$. 
	The phase derived for this case is shown in Fig.~\ref{fig:phase_caso2} (full line). Since the phase behaves differently, in different regions of the atmospheric layers, we will analyze the outermost layers and innermost layers separately. 
	
\begin{figure}
\begin{center}
\includegraphics[width=20pc]{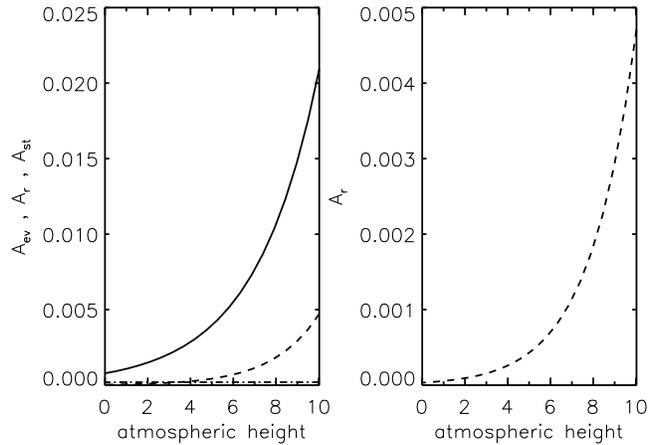}
\caption{Case 2 - same as in figure~\ref{fig:phase_caso2}.
Left panel: Dimensionless amplitudes derived from each term of the velocity: full line - A$_{ev}$; dashed line - A$_{r}$; dashed-dotted line - A$_{st}$
Right panel: Zoom in the y axis for A$_{r}$.}
\label{fig:amp_sob_caso2}
\end{center}
\end{figure}

 Higher in the atmosphere, in the region where the phase decreases with atmospheric height, the amplitude derived from the evanescent term alone is again much larger than the amplitude derived from the running or magnetic terms alone, as can be seen in Fig.
 \ref{fig:amp_sob_caso2}.\
 
\begin{figure}
\begin{center}
\includegraphics[width=20pc]{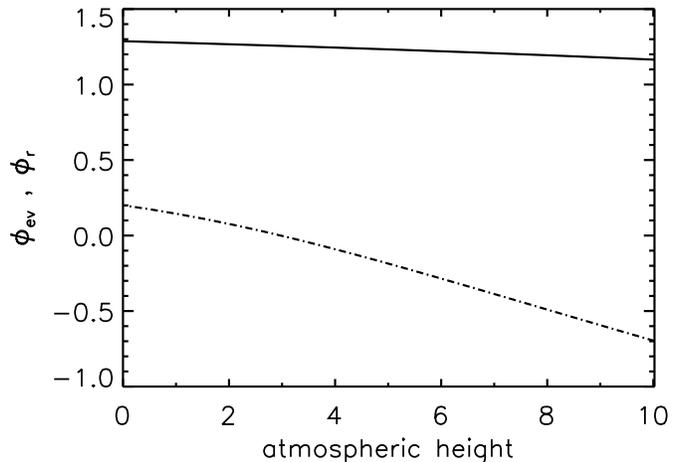}
\caption{Case 2 - same as in figure~\ref{fig:phase_caso2}. Phase derived from the evanescent term alone (full line) and from the running term alone (dashed-dotted line) as a function of atmospheric height (in units of $\mathcal{H}$).}
\label{fig:phase_ev_caso2}
\end{center}
\end{figure}

Thus, in the outermost layers this case seems similar to case study 1, the main difference being that in the latter the total phase was constant, while here it decreases slightly with atmospheric height. To understand why the phase varies with atmospheric height, we will first consider the phase of the dominating term, namely, $\phi_{ev}$, which is shown on Fig.~\ref{fig:phase_ev_caso2} (full line). 

Unlike what was found in case study 1, in the present case $\phi_{ev}$ varies with atmospheric height. To check that we understand the reason for this behaviour, we turn to the toy study presented in section \ref{toy model}. In our toy study the analogy with this case is achieved by considering two evanescent waves. The amplitude of the first wave, $\psi_1$, is considered to be larger than the amplitude of the second wave, $\psi_2$, inspired in the fact that the amplitude of the acoustic component of the velocity decreases with colatitude. Moreover, the acoustic wave number of the first wave, $k_1$, is considered to be larger than the acoustic wave number of the second wave, $k_2$, inspired in the fact that the absolute value of the acoustic vertical wavenumber for the evanescent term decreases with colatitude. 

\begin{figure}
\begin{center}
\includegraphics[width=18pc]{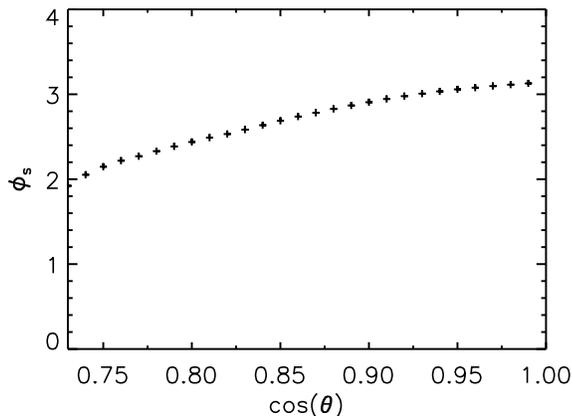}
\caption{Case 2 - same as in figure~\ref{fig:phase_caso2}. Zoom of the input phase of the acoustic term, $\phi_s$, in the region where the solution is evanescent as function of colatitude. }
\label{fig:phase_input_s_c2}
\end{center}
\end{figure}

The contributions of the two waves to the numerator and denominator of equation (\ref{eq: 2.1.7}) depend critically on the phases $\phi_1$ and $\phi_2$. These can be estimated from the input phases of our real model. In Fig.~\ref{fig:phase_input_s_c2} we show the input phase of the acoustic component, $\phi_s$, in the region of colatitude where the acoustic component is evanescent. We conclude from the plot that the value of $\phi_1$ (the wave nearest the magnetic pole) is such that $|\cos(\phi_1)|$ is close to 1 and thus larger than $|\cos(\phi_2)|$. Since the amplitude, $\psi_1$, is also larger than $\psi_2$ we are going to assume that the contribution of the second wave to the numerator of equation (\ref{eq: 2.1.7}) can be neglected, in which case the latter becomes, 

\begin{equation}
\label{eq: ev_desce_c2}
\phi_{toy}=\rm{arccot}(-\tan(\phi_1)-\frac{\psi_2}{\psi_1}e^{(k_1-k_2)\eta}\frac{\sin(\phi_2)}{\cos(\phi_1)}).
\end{equation}

 The behaviour of $\phi_{toy}$ will therefore depend critically on the sign of $\cos(\phi_1)$ and $\sin(\phi_2)$. From Fig. \ref{fig:phase_input_s_c2} we see that $\cos(\phi_s)$ is always negative and, thus, so is $\cos(\phi_1)$, and $\sin(\phi_s)$ is always positive and, thus, so is $\sin(\phi_2)$. Therefore, since $e^{{(k_1-k_2)}\eta}$ increases with atmospheric height, $\phi_{toy}$ will decrease with atmospheric height. We note that such conclusion would be reached if the term of $\phi_2$ in the denominator had not been neglected, but the algebra would be more complicated.\

Considering the results of our model, and equation (\ref{eq: ev_desce_c2}), we may be tempted to conclude that the phase behavior seen in the outermost layers in this case study is due to a visual superposition of evanescent waves of different wave numbers and input phases, with the resulting phase decreasing with atmospheric height. However, when comparing the phase derived from the evanescent term alone, (full line in Fig.\ref{fig:phase_ev_caso2}), with the total phase shown in Fig. \ref{fig:phase_caso2} higher in the atmosphere, we find that their slopes are not the same. Therefore, one of the other two components of the velocity must be contributing to the behavior of the total phase. In fact, looking at Fig.~\ref{fig:phase_caso2} where the phase derived from the acoustic term alone is shown (dashed-dotted line), one can see that the slope of the latter is very similar to the slope of the total phase in the outermost layers. Thus we may conclude that it is the acoustic running component, whose phase also decreases with atmospheric height (cf.~Fig.\ref{fig:phase_ev_caso2}), that is influencing, together with the evanescent component, the behavior of the phase in the outermost layers of the atmosphere. This is an example of a situation in which a component of significant lower amplitude influences the final result.

\begin{figure}
\begin{center}
\includegraphics[width=18pc]{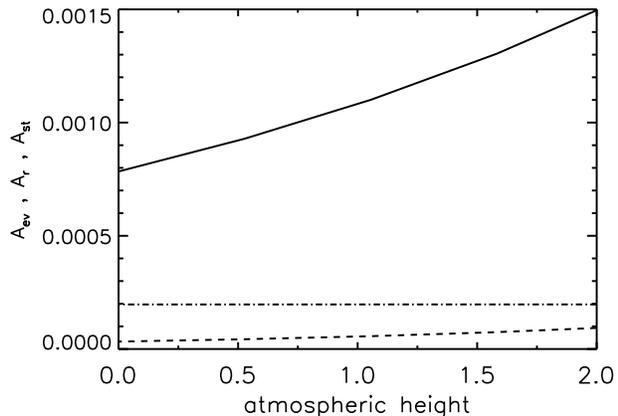}
\caption{Case 2 - same as in figure~\ref{fig:phase_caso2}. Zoom in the xx axis for the dimensionless amplitudes derived from each term of the velocity: full line - A$_{ev}$; dashed line - A$_{r}$; dashed-dotted line - A$_{st}$. }
\label{fig:amp_[0-2]_caso2}
\end{center}
\end{figure}

Deeper in the atmosphere, in the region where the phase is increasing with atmospheric height, the amplitude derived from the evanescent term alone is still larger than the amplitude derived from the running and magnetic terms alone, as can be seen in Fig. \ref{fig:amp_[0-2]_caso2}. Thus, we could naively expect the evanescent term to dominate the phase behavior. However, from a comparison of Figs~\ref{fig:phase_caso2} and \ref{fig:phase_ev_caso2}, we see that such is clearly not the case. 

\begin{figure}
\begin{center}
\includegraphics[width=18pc]{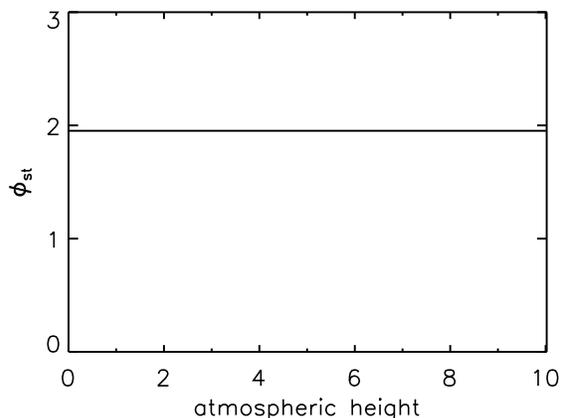}
\caption{Case 2 - same as in figure~\ref{fig:phase_caso2}. Phase derived from the magnetic term alone as a function of atmospheric height (in units of $\mathcal{H}$). }
\label{fig:phase_st_caso2}
\end{center}
\end{figure}

Looking back at our toy study, the analogy with the case presented here is achieved by considering the first wave to be an evanescent wave, the second wave to be a running wave, and the third wave to be a standing wave with constant amplitude.
The amplitudes of the wave-like functions in the toy model are estimated from the amplitudes derived when considering the evanescent, the running, and the standing terms alone. Similarly, the phases in the toy model are estimated from the phases of the corresponding components, adding $\pi/2$, i.e., $\phi_1=\phi_{ev}+\pi/2$, $\phi_2=\phi_{r}+\pi/2$, and $\phi_3=\phi_{st}+\pi/2$.  The $\pi/2$ difference is a consequence of using the sin function in equation~(\ref{v_1}) and the cos function in the fits to extract $\phi_{ev}$, $\phi_{r}$, and $\phi_{st}$.

Due to its amplitude, we find, after checking the terms in equation (\ref{eq: phi_3}), that the contribution of the running wave to $\phi_{toy}$ can be neglected. Moreover, from the inspection of $\phi_{st}$ we find that we can also neglect the contribution of the standing wave to the denominator of equation (\ref{eq: phi_3}). However, the contribution of the latter to the numerator of equation (\ref{eq: phi_3}) cannot be neglected, as happened higher in the atmosphere. In fact, the amplitude of the evanescent wave deeper in the atmosphere is much smaller than its amplitude higher in the atmosphere, and when the phases of the evanescent and standing components are taken into account, the corresponding terms in the numerator of equation (\ref{eq: phi_3}) become of the same order of magnitude. Then equation (\ref{eq: phi_3}) becomes, 

\begin{equation}
\label{eq: phase_sobe}
\phi_{toy}=\arctan(-\cot(\phi_1)-\frac{\psi_3}{\psi_1}e^{\eta/2\mathcal{H}}e^{-k_1\eta}\frac{\cos(\phi_3)}{\sin(\phi_1)}).
\end{equation}

The behaviour of $\phi_{toy}$ will, therefore, depend critically on the sign of $\cos(\phi_3)$ and $\sin(\phi_1)$. From Fig.\ref{fig:phase_st_caso2} and the relation $\phi_3= \phi_{st}+\pi/2$, one concludes that $\cos(\phi_3)$ is negative, and from Fig.~\ref{fig:phase_ev_caso2} and the relation $\phi_1= \phi_{ev}+\pi/2$, one concludes that $\sin(\phi_1)$ is positive. Since, $e^{\eta/2\mathcal{H}}e^{-k_1\eta}$ increases with atmospheric height, and $\phi_1$ varies only very little within those lower layers, the result will be a phase increasing with atmospheric height. \

Considering the results of this case study, and equation~(\ref{eq: phase_sobe}), we may thus come to the conclusion that the phase behavior seen in the deeper layers is likely due to a visual superposition of the evanescent acoustic waves and magnetic waves, with the resulting phase increasing with atmospheric height.

\vspace{0.5cm}

\underline{Case 3: Phase increasing with atmospheric height}

\vspace{0.5cm}

Case study 3 is for $B_p$=1~kG, $\nu=1.7$~mHz, and an integration over the whole visible stellar disk.  The oscillation frequency in this case is smaller than the acoustic cutoff frequency. The critical angle is 67\arcdeg.

\begin{figure}
\begin{center}
\includegraphics[width=18pc]{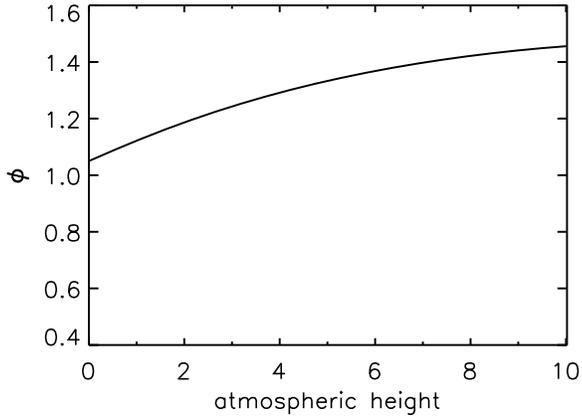}
\caption{Case 3 - $B_{p}$=1 kG,\ $\nu=1.7$ mHz, and an integration over the visible stellar disk. Total phase as a function of atmospheric height(in units of $\mathcal{H}$).}
\label{fig:phase_total_caso3}
\end{center}
\end{figure}

\begin{figure}

\begin{center}
\includegraphics[width=20pc]{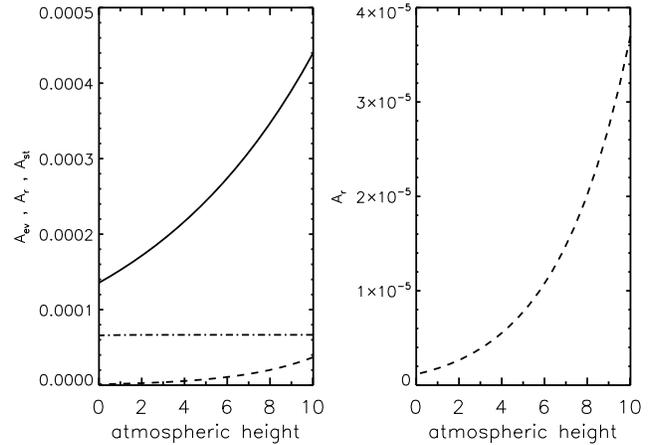}
\caption{Case 3 - same as in figure~\ref{fig:phase_total_caso3}.
Left panel: Dimensionless amplitudes derived from each term of the velocity: full line - A$_{ev}$; dashed line - A$_{r}$; dashed-dotted line - A$_{st}$.
Right panel: Zoom in the y axis for A$_{r}$.}
\label{fig:amp_sob_caso3}
\end{center}
\end{figure}

From Fig.\ref{fig:phase_total_caso3} we see that in this case the phase obtained with equation (\ref{eq: 2.59}) is increasing throughout the region of the atmosphere under consideration. In Fig. \ref{fig:amp_sob_caso3} we show the amplitude derived from each of the components of the velocity alone. As before,  we find that the amplitude derived from the evanescent term alone is always much larger than the amplitude derived from the running or magnetic terms alone. Yet, the behaviour of the phase is different from that of the case studies considered before.

\begin{figure}
\begin{center}
\includegraphics[width=18pc]{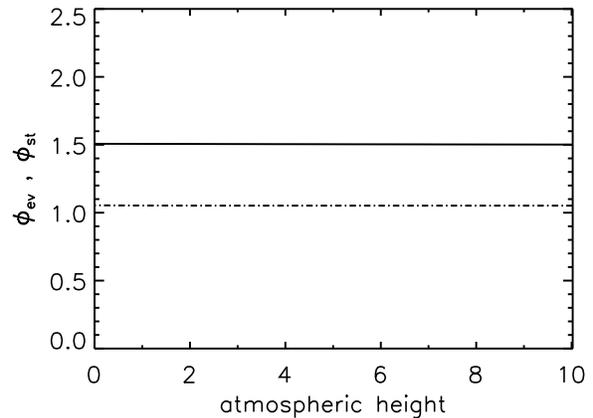}
\caption{Case 3 - same as in figure~\ref{fig:phase_total_caso3}. Phase derived from the evanescent term alone (full line) and phase derived from the magnetic term alone (dashed-dotted line) as a function of atmospheric height (in units of $\mathcal{H}$).}
\label{fig:phase_ev_c3}
\end{center}
\end{figure}

To understand the phase behaviour, we again look for the analogous case in our toy model, taking the first, second, and third waves, as evanescent, running, and standing waves, respectively.

Due to its amplitude, we find after checking the magnitude of the terms in equation (\ref{eq: phi_3}), that the contribution of the running wave to $\phi_{toy}$ can again be neglected. Moreover, looking at Fig.\ref{fig:phase_ev_c3}, where we show the phases derived from the standing and evanescent terms alone, respectively, and taking into account that $\phi_3=\phi_{st}+\pi/2$ and $\phi_1=\phi_{ev}+\pi/2$, we see that we can neglect the contribution of the standing wave to the numerator of equation (\ref{eq: phi_3}). However, the contribution of the latter to the denominator of equation (\ref{eq: phi_3}) cannot be neglected, since $\sin(\phi_1)$ is sufficiently small that the evanescent and standing contributions become of the same order. Then equation (\ref{eq: phi_3}) becomes, 

\begin{equation}
\label{eq: phase_sobe_c3}
\phi_{toy}=\rm{arccot}(-\tan(\phi_1)-\frac{\psi_3}{\psi_1}e^{-\eta/2\mathcal{H}}e^{k_1\eta}\frac{\sin(\phi_3)}{\cos(\phi_1)}).
\end{equation}

From Fig. \ref{fig:phase_ev_c3}, one infers that $\sin(\phi_3)$ is positive and $\cos(\phi_1)$ is negative. Since, $e^{-\eta/2\mathcal{H}}e^{k_1\eta}$ decreases with atmospheric height and $\tan(\phi_1)$ varies only very little within the atmospheric region considered, the result will be a phase increasing with atmospheric height.\

From equation (\ref{eq: phase_sobe_c3}) we may thus conclude that the phase behavior seen in this case study is likely due to a visual superposition of the evanescent and magnetic terms, with the resulting phase increasing with atmospheric height.

We note that there are other situations that can result in the phase increasing with atmospheric height. Such behavior could be found as a result of a visual superposition of running and evanescent waves, or of running and standing waves, under particular conditions.

\vspace{0.5cm} 
 
\underline{Case 4: Phase decreasing deeper and constant higher}
\underline{in the atmosphere}

\vspace{0.5cm}

The present case is for $B_p$=1~kG and $\nu=2.1$ mHz. The integration is over the whole visible stellar disk. The oscillation frequency is smaller than the acoustic cutoff frequency and the critical angle is 57$\arcdeg$. As in case study 2, in this case the phase behaves differently, in different atmospheric layers. Thus, we will analyze the outermost layers and innermost layers separately.   

\begin{figure}
\begin{center}
\includegraphics[width=18pc]{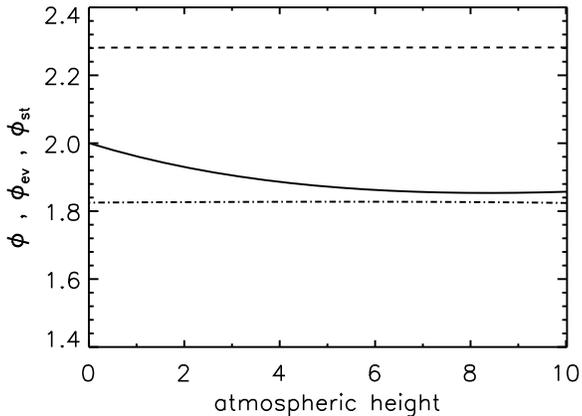}
\caption{Case 4 - $B_{p}$=1 kG,\ $\nu=2.1$ mHz, and an integration over the visible stellar disk. Total phase (full line), phase derived from the evanescent term alone (dashed-dotted line) and phase derived from the magnetic term alone (dashed line) as function of atmospheric height (in units of $\mathcal{H}$).}
\label{fig:phase_total_caso4.1}
\end{center}
\end{figure}

\begin{figure}
\begin{center}
\includegraphics[width=20pc]{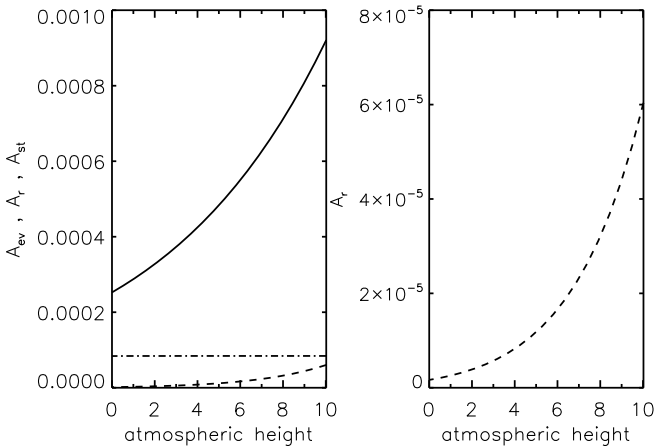}
\caption{Case 4 - same as in figure~\ref{fig:phase_total_caso4.1}.
Left panel: Dimensionless amplitudes derived from each term of the velocity: full line - A$_{ev}$; dashed line - A$_{r}$; dashed-dotted line - A$_{st}$
Right panel: Zoom in the y axis for A$_{r}$.}
\label{amp_sob_c4}
\end{center}
\end{figure}

Higher in the atmosphere, in the region where the phase is approximately independent of the atmospheric height, the amplitude derived from the evanescent term alone is again significantly larger than the amplitude derived from the running or magnetic terms alone, as we can see in Fig.\ref{amp_sob_c4}. Moreover, if we compare the evanescent phase shown in Fig.\ref{fig:phase_total_caso4.1} (dashed-dotted line), with the total phase in the same figure (full line), we may conclude that the result is consistent with the evanescent term dominating the phase behaviour in that region of the atmosphere. The difference between the values of the phase can be explained by the influence of the lower atmospheric layers on $\phi$.

Deeper in the atmosphere, in the region where the phase is decreasing with atmospheric height, the amplitude derived from the evanescent term alone is still larger than the amplitudes derived from the running and magnetic terms alone. 
 However, looking at the phase derived from the evanescent term alone, we can see that it does not show the same behavior, deeper in the atmosphere, as the total phase. Thus, we may conclude that at least one of the other terms must be influencing the behavior of the total phase.   

%\begin{figure}
%\begin{center}
%\includegraphics[width=18pc]{amp_sobre_[0-2]_c4}
%\caption{Case 4 - observer at $0$ \arcdeg, $B_{p}$=1kG,\ $\nu=2.13$ mHz, \textit{l}=1 and for an integration over the visible stellar disk. Zoom in the xx axis for the dimensionless amplitude derived from each term of the velocity alone as function of atmospheric height (in units of $\mathcal{H}$): full line - A$_{ev}$; dashed line - A$_{r}$; dashed-dotted line - A$_{st}$. }
%\label{fig:amp_[0-2]_caso4}
%\end{center}
%\end{figure}

To investigate further, we look back at our toy study, identifying, as before, the first, second, and third waves with, evanescent, running, and standing waves, respectively.
Due to its amplitude and phase, the contribution of the running wave to $\phi_{toy}$ can be neglected. Moreover, from Fig.~\ref{fig:phase_total_caso4.1}, we see that $\cos(\phi_1)\sim \cos(\phi_{ev}+\pi/2)$ will not be small and, thus, we may neglect the contribution of the standing wave to the numerator of equation (\ref{eq: phi_3}). However, the contribution of the latter to the denominator of equation (\ref{eq: phi_3}) cannot be neglected, as happened higher in the atmosphere. As seen in Fig. \ref{amp_sob_c4}, the amplitude of the evanescent wave deeper in the atmosphere is considerably smaller than its amplitude higher in the atmosphere, and when the phases of the evanescent and standing components are taken into account the terms in the denominator of equation (\ref{eq: phi_3}) become of the same order of magnitude. Then equation (\ref{eq: phi_3}) takes the form, 

\begin{equation}
\label{phase_desce_c4}
\phi_{toy}=\rm{arccot}(-\tan(\phi_1)-\frac{\psi_3}{\psi_1}e^{-\eta/2\mathcal{H}}e^{k_1\eta}\frac{\sin(\phi_3)}{\cos(\phi_1)}).
\end{equation}

This equation is the same as equation (\ref{eq: phase_sobe_c3}). However, in this case we deduce from Fig.\ref{fig:phase_total_caso4.1}  that $\sin(\phi_3)$ is negative and from Fig.\ref{fig:phase_total_caso4.1} that $\cos(\phi_1)$ is also negative. Thus, in this case we find a phase decreasing with atmospheric height.\
 
From equation (\ref{phase_desce_c4}), we may thus conclude that the phase behavior seen in the deeper layers in this case study is likely due also to a visual superposition of the evanescent and magnetic terms, but in this case the resulting phase decreases with atmospheric height.

\vspace{0.5cm}

\underline{Case 5: Phase decreasing fast throughout the whole}
\underline{atmospheric layer}

\vspace{0.5cm}

This case study is for $B_p=2$~kG, $\nu=3.0$ mHz, and an integration over the whole visible stellar disk. The oscillation frequency is larger than the acoustic cutoff frequency, therefore we do not have a critical angle: the acoustic term will be a running wave at all latitudes. 

\begin{figure}
\begin{center}
\includegraphics[width=18pc]{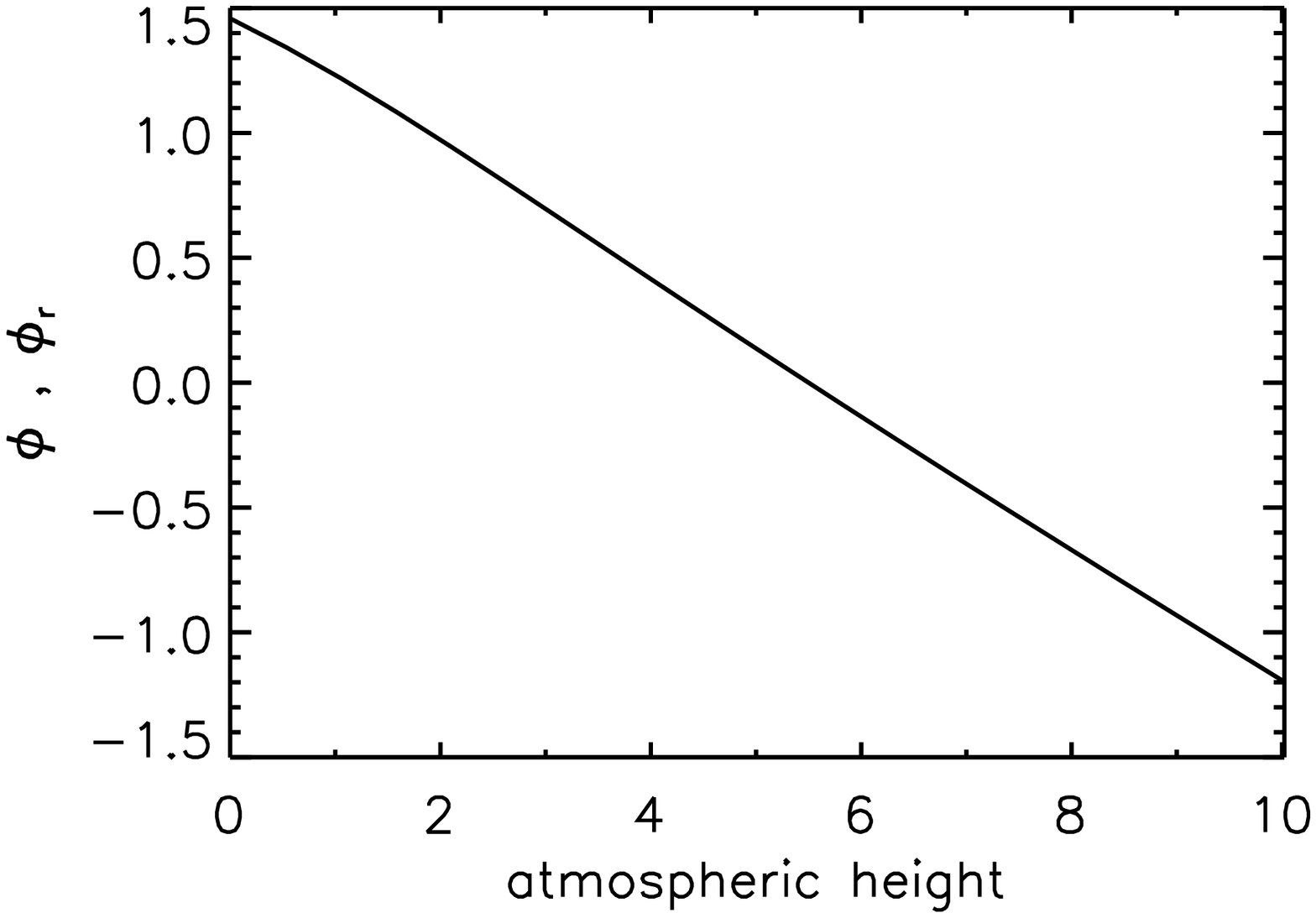}
\caption{Case 5 - $B_{p}$=2 kG, $\nu=3.0$ mHz, and an integration over the visible stellar disk. Total phase overlaped with phase derived from the running term alone as a function of atmospheric height (in units of $\mathcal{H}$). The two lines are coincident.}
\label{fig:phase_c5}
\end{center}
\end{figure}

\begin{figure}
\begin{center}
\includegraphics[width=18pc]{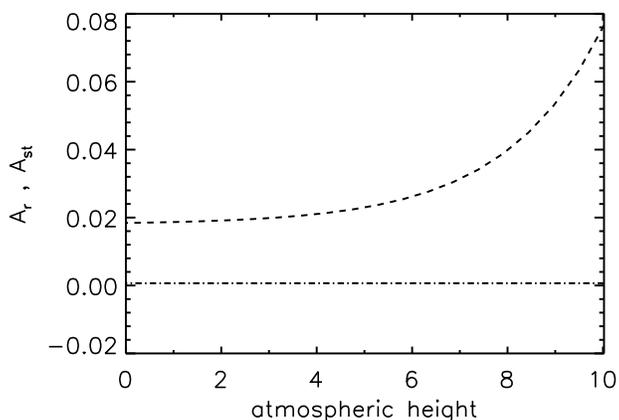}
\caption{Case 5 - same as in figure~\ref{fig:phase_c5}. Dimensionless amplitudes derived from each term of the velocity: dashed line - A$_{r}$; dashed-dotted line - A$_{st}$}.
\label{fig:amp_sob_c5}
\end{center}
\end{figure}

From Fig.~\ref{fig:phase_c5}, we see that in this case the phase obtained with equation (\ref{eq: 2.59}) is decreasing throughout the whole region of the atmosphere under consideration. In Fig.~\ref{fig:amp_sob_c5} we show the amplitude derived from each of the components of the velocity alone. Clearly, the amplitude derived from the running term alone is always much larger than the amplitude derived from the magnetic term alone. 

Moreover, Fig.~\ref{fig:phase_c5}, shows that the phase derived from the running term alone overlaps with the total phase, and we may conclude that the running term dominates also the phase behavior. This is a again a simple case, in which one of the terms of the velocity dominates entirely the final result.

Nevertheless, it is important to notice, as before, that other situations can exist in which the resulting phase decreases with atmospheric height due to reasons different than the one considered in this case study. Whatever the visual superposition, if it involves running waves that, due to their amplitude and phase, are dominating, the result will be a phase decreasing with atmospheric height.  A phase decreasing with atmospheric height can also occur due to a visual superposition of running and evanescent waves, of running and standing waves, or even of evanescent waves with slightly different wave numbers. Despite this, we note that when the resulting phase is decreasing with atmospheric height due to a visual superposition of waves of different nature, the rate at which it decreases is significantly smaller than when it is due to the running waves alone.  The only exception to this is the case when the pulsation amplitude approaches a zero, and the phase decreases significantly, in some limited region of the atmosphere. That is the situation that will be investigated in case study 6, below.

\vspace{0.5cm}

\underline{Case 6: A node in the amplitude and a phase jump}
\vspace{0.5cm}

The last case study to be considered is for $B_p$=4~kG, $\nu=2.2$~mHz, and an integration over colatitudes within the interval [62\arcdeg,85\arcdeg].   
The oscillation frequency is smaller than the acoustic cutoff frequency and the critical angle is 55.5\arcdeg. Since the integration is over a colatitude region that is above the critical angle, the acoustic term has always the form of a running wave, with a characteristic wavelength and amplitude which depend on co-latitude. 

\begin{figure}
\begin{center}
\includegraphics[width=18pc]{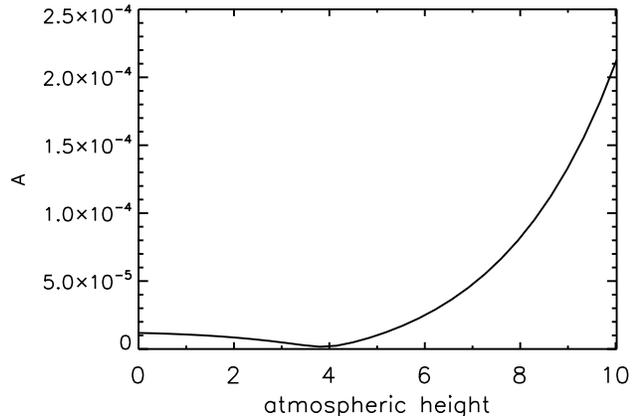}
\caption{Case 6 - $B_{p}$=4 kG,\ $\nu=2.2$ mHz, and an integration of [62,85]$\arcdeg$ of colatitude. Dimensionless amplitude as a function of  atmospheric height (in units of $\mathcal{H}$). }  
\label{fig:amplitude_nodo}
\end{center}
\end{figure}

\begin{figure}
\begin{center}
\includegraphics[width=18pc]{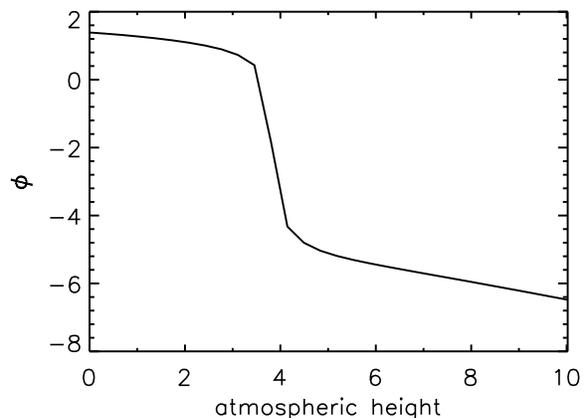}
\caption{Case 6 - same as in figure~\ref{fig:amplitude_nodo}. Phase as a function of atmospheric height (in units of $\mathcal{H}$). } 
\label{fig:phase_nodo}
\end{center}
\end{figure}

In Figs. \ref{fig:amplitude_nodo} and \ref{fig:phase_nodo}, where the results for this case study are shown, we can see a zero in the amplitude accompanied by a phase jump of $\sim\pi$, at an atmospheric height close to 4$\mathcal{H}$. Although at first sight this behaviour may be thought to result from a node in the velocity eigenfunction, we know that such cannot be the case. In fact, from our study of the velocity field we know that in this region of our model atmosphere the component of the velocity along the magnetic field is, at each latitude, a running wave, and the component perpendicular to the magnetic field has an amplitude very closely independent of atmospheric height. Thus, the velocity eigenfunction cannot have the properties of a standing wave with a node in the atmospheric layers considered. We can, therefore, anticipate that the behavior seen in the amplitude and phase in this case study is only that of a ``false node", that results from the visual superposition of the line-of-sight projections of the acoustic and magnetic components, when integrated over the stellar region considered.

\begin{figure}
\begin{center}
\includegraphics[width=18pc]{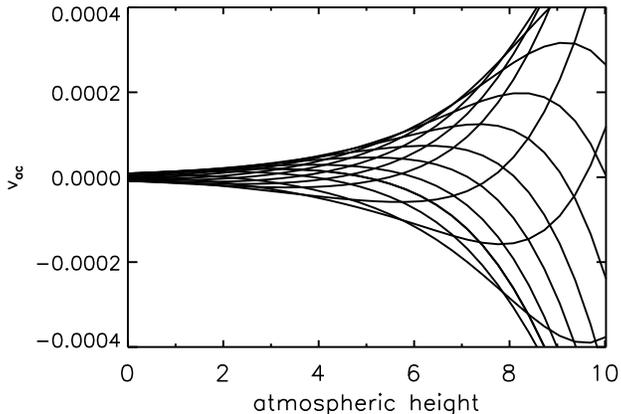}
\caption{Case 6 - same as in figure~\ref{fig:amplitude_nodo}. Dimensionless acoustic component of the integrated line-of-sight velocity as a function of atmospheric height (in units of $\mathcal{H}$). Different lines are for different instants during the pulsation cycle.}  
\label{fig:v_ac}
\end{center}
\end{figure}

\begin{figure}
\begin{center}
\includegraphics[width=18pc]{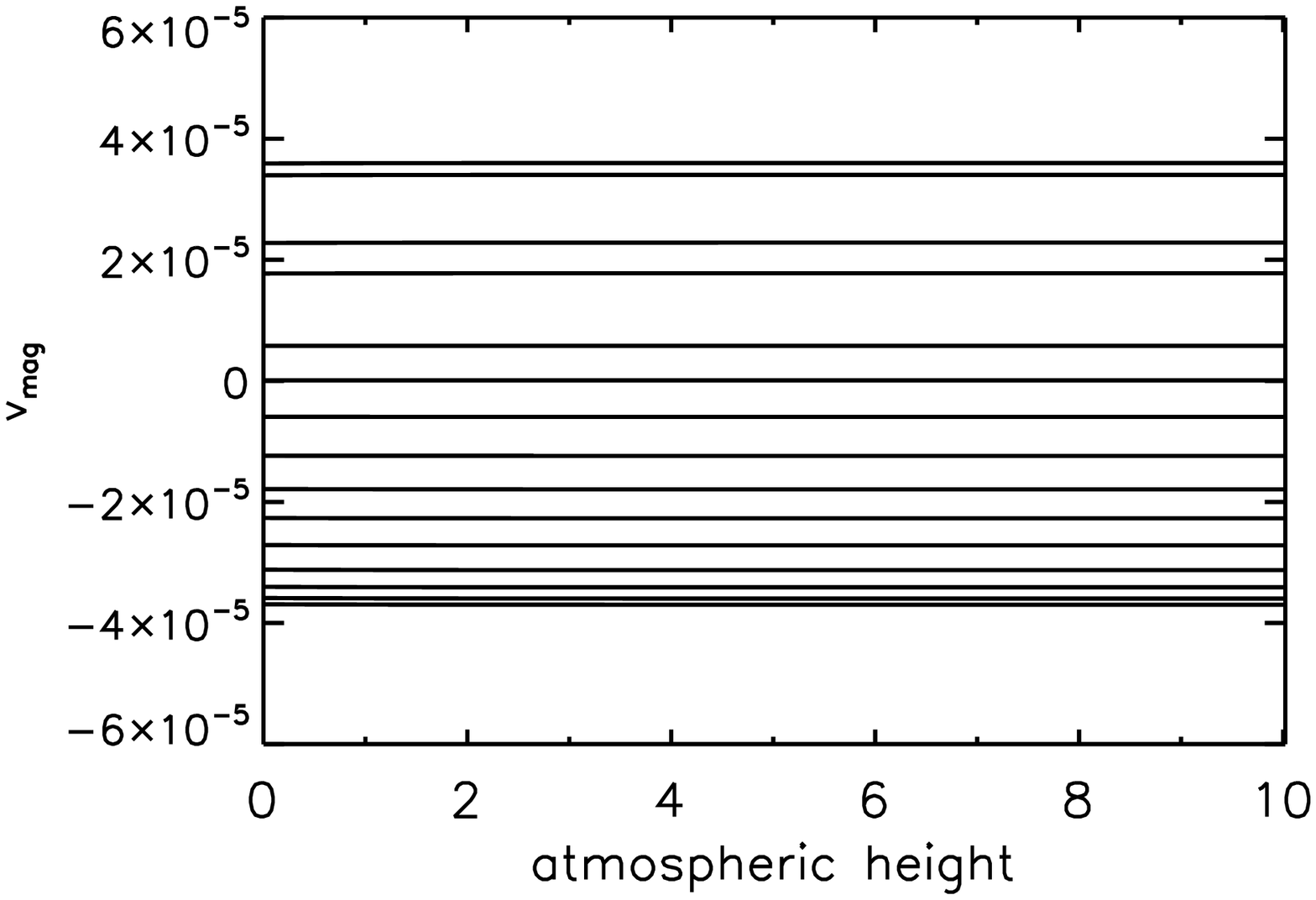}
\caption{Case 6 - same as in figure~\ref{fig:amplitude_nodo}. Dimensionless magnetic component of the integrated line-of-sight velocity as a function of atmospheric height (in units of $\mathcal{H}$). Different lines are for different instants during the pulsation cycle.}  
\label{fig:v_mag}
\end{center}
\end{figure}

\begin{figure}
\begin{center}
\includegraphics[width=18pc]{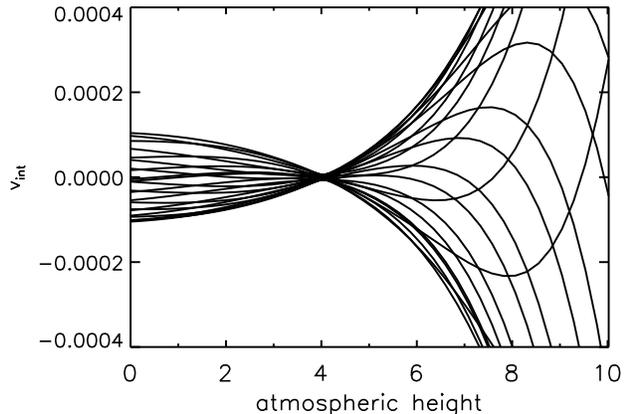}
\caption{Case 6 - same as in figure~\ref{fig:amplitude_nodo}. Dimensionless integrated line-of-sight velocity as a function of atmospheric height (in units of $\mathcal{H}$). Different lines are for different instants during the pulsation cycle.}  
\label{fig:v_total}
\end{center}
\end{figure}

\begin{figure}
\begin{center}
\includegraphics[width=18pc]{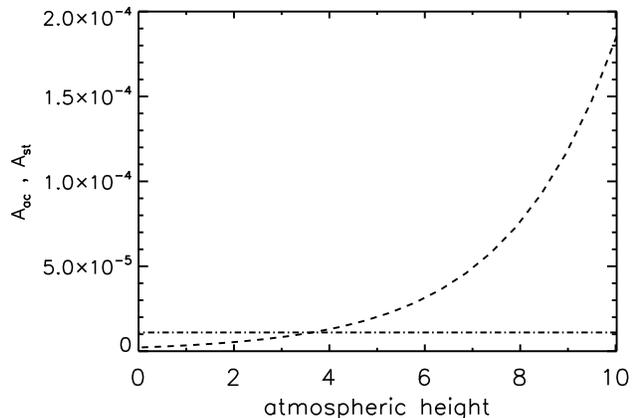}
\caption{Case 6 - same as in figure~\ref{fig:amplitude_nodo}. Dimensionless amplitudes derived from each term of the velocity: dashed line - $A_{ac}$; dashed-dotted line - $A_{st}$.}  
\label{fig:amp_sob_caso7}
\end{center}
\end{figure}

In Figs.~\ref{fig:v_ac} and \ref{fig:v_mag} we show, respectively, the acoustic, $v_{ac}$, and magnetic, $v_{mag}$, components of the integrated line-of-sight velocity, $v_{int}$. We note that in the present case $v_{mag}$ is identical to the integral $I_3$, defined in equation~(\ref{standing}), while $v_{ac}$ is a subsection of the integral $I_2$, corresponding to the interval of colatitudes under consideration. Different lines correspond to different times, during a pulsation cycle. It is clear that none of the integrated velocity components shows the characteristics of a standing wave with a node. Nevertheless, the signature of the apparent node is evident in $v_{int}$, shown in Fig.~\ref{fig:v_total}. From the inspection of Fig. \ref{fig:amp_sob_caso7}, where we show the amplitudes derived from each of the integrated velocity components, we see that the apparent node results from the cancellation between the two terms of the integral for the total line-of-sight integrated velocity, when averaged over the region of the stellar disk considered. The jump in the total phase results from the change in the sign of $v_{int}$, like it would occur when crossing a node in a pure standing wave.

%Looking now back at Fig.\ref{fig:amplitude_nodo}, we can see that high in the atmosphere the amplitude is increasing. This is due to the fact that higher in the atmosphere the amplitude derived for the acoustic term alone is always much larger than the amplitude derived for the magnetic term alone, because the former increases exponentially with atmospheric height. Deeper in the atmosphere the two amplitudes become of the same order and at $\mathcal{H}$=4 scale heights they cancel out, as mentioned before.  
%At this atmospheric height, the dependence of $\eta$ in the amplitude expression for the acoustic term passes through a minimum, making the amplitude of this term similar to the amplitude of the magnetic term, resulting in an apparent node on the amplitude plot. Higher in the atmosphere, although the dependence of $\eta$ remains low, it is compensated by the variation of the atmospheric pressure.\

%Accompanying the apparent node in the amplitude we see the jump in the total phase, and far from the apparent node one can see a phase decreasing with atmospheric height. Like previously mentioned, in this case, the acoustic term will be a running wave and thus we can foresee that this pattern seen far from the node must be due to a visual superposition of running wave and standing waves, with the resulting phase decreasing with atmospheric height. 

\section{Discussion}
\label{conclusions}

The results presented in the previous section are to be considered as illustrative of what may take place in the outer layers of the atmosphere, hence, in the region where spectral lines associated with rare earth elements are formed. In deeper layers of the atmosphere, the assumed isothermal condition becomes inappropriate and, depending on the magnetic field intensity, the magnetoacoustic wave may not yet be decoupled into acoustic and magnetic components.

If rare earths are indeed formed in the region of the atmosphere considered in this work, then a phase jump of $\pi$ associated with a zero of the pulsation amplitude, derived from these lines, cannot be due to a real pulsation node. Nevertheless, the same behaviour of the phase and amplitude derived from lines that are sensitive to deeper regions of the atmosphere (say, the base of the photosphere) may still reflect the presence of a real pulsation node. The analysis of the theoretical radial velocity near the base of the photosphere is beyond the scope of this paper, thus we cannot make conclusions with regards to alleged nodes derived from, e.g, the inspection of the core of H $\alpha$ lines, or of the Fe and Mg lines \citep{2005MNRAS.364..864E}. However, there are reasons to believe that real nodes may exist at the photospheric level, at least in some latitudinal regions of the star \citep{1998MNRAS.301...31G,2009ApJ...704.1218K,2010MNRAS.403.1729S}, and result in radial velocity amplitude and phase behaviours that are similar to those of false nodes. 

Despite the diversity of phase behaviour found in the results presented in sec.~\ref{results}, it is clear that in the generality of the case studies considered the derived phase variations are rather small. In fact, some of these are unlikely to be detected, given the uncertainties in the phases derived from the time-series of spectra. Large and rapid phase variations are derived in case studies 5 and 6. The first of these concerns a situation in which the running wave largely dominates the integral for the radial velocity, while the second is the case of the fake node where the phase varies at a high rate in a localized region of the atmosphere. Additional situations of strong variations of the phase, as well as of false nodes, are found when different positions of the observer and mode degrees are considered. Those shall be detailed in paper~II.  

Looking more generally at our results, we may conclude that the behaviour of the theoretically derived amplitudes is in good qualitative agreement with that inferred from the observations, the most common trend being an increase with atmospheric height. 

Concerning the pulsation phases, our findings illustrate well the difficulties that may arise when attempting to infer information about the properties of the pulsations in the atmospheres of roAp stars through the inspection of the spectroscopic time-series. Those difficulties are due, in particular, to the observer's inability to resolve the star. The observations need to be thought of as visual superposition of different wave functions, and / or similar wave functions with different characteristic wavelengths, amplitudes, and phases, which are not always easy to disentangle. A clear example of this is the fact that the same general phase behaviour can  be explained by combinations of different wave-type solutions. 

In fact, phases that are independent of atmospheric height may indicate that evanescent waves of acoustic nature, or standing waves of magnetic nature, dominate  the integral for the radial velocity, or that both type of waves contribute similarly to the latter. Similarly, different situations can lead to an increase of the phase with increasing atmospheric height, such as the visual superposition of evanescent and standing waves, the visual superposition of running and evanescent waves, or the visual superposition of running and standing waves. Moreover, this behaviour of the phase can also result simply from a visual superposition of evanescent waves with slightly different wave numbers. We should note, however, that for lines formed in the region of the atmosphere considered in the present work, this behaviour of the phase cannot be due to a downwardly propagating wave, since no such type of solution is present there. Finally,  phases decreasing with atmospheric height can result from running waves propagating upwardly, from a visual superposition of running and evanescent waves, or of running and standing waves, and from a visual superposition of evanescent waves with slightly different wave numbers.

Despite the potential difficulties that may arise when attempting to disentangle the different possible scenarios, it is also clear that some of these lead to substantially different rates of phase changes, as well as amplitude changes with atmospheric height. Thus, it is likely that at least some of the apparent degeneracy in phase behaviour may disappear if a more realistic model of the stellar atmosphere,  than that used in the present analysis, is considered, and a direct confrontation of observations and theoretical predictions is carried out.

    Finally, as acknowledged before, there are other aspects of the parameter space that must be taken into consideration, like the position of the observer, the mode degree and the oscillation frequency. Any definitive conclusion about a given observed phenomenon requires that the whole parameter space is explored. Results of such an analysis are to be presented in paper~II.

%\bsp

\label{lastpage}
\section*{acknowledgements}
 This work was supported by the project PTDC/CTE-AST/098754/2008 and the grant SFRH/BD/38908/2007 funded by FCT/MCTES, Portugal. MC is supported by a Ci\^encia 2007 contract, funded by FCT/MCTES(Portugal) and POPH/FSE (EC).

\bibliographystyle{mn2e}
\bibliography{referencias}
APPENDIX A: ORDER OF MAGNITUDE ANALYSIS
\label{app-A1}

\vspace{0.3cm}
\underline{\textbf{Slow Acoustic Component}}
\vspace{0.3cm}

To derive equations~(\ref{eq: 2.26}) and (\ref{eq: 2.25}), we have considered the order of magnitude of each term in equations~(\ref{eq: 2.18}) and (\ref{eq: 2.19}), after separating the time and the slow and fast components of the displacement. For that we will take as a reference the solution that we would obtain in the absence of the magnetic field. The latter is of the type,

\begin{equation}
\label{eq: 2.22b}
{\xi}=\frac{A}{p^{1/2}}e^{-ikz}+\frac{B}{p^{1/2}}e^{ikz},
\end{equation}
with $A$ and $B$ being constant amplitudes and, 

\begin{equation}
\label{eq: 2.22b}
k^2=\frac{\omega^2-\omega^2_c}{c^2},
\end{equation}
where $\omega^2_c$ is the acoustic cutoff frequency defined to be $\omega_c=c/(2H)$ and $c^2={\gamma p_0}/{\rho_0}$ is the sound speed in an isothermal atmosphere \citet{1993afd..conf..399G}. In the above, $A=0$ if $\omega<\omega_c$ and $B=0$ if $\omega>\omega_c$. 

We may expect that the solution for the slow acoustic component of the displacement will vary on a scale similar to the solution in the absence of the magnetic field. Based on this assumption and considering first the case in which $B_x \approx B_z$, we can study the order of magnitude of the terms in equation (\ref{eq: 2.19}). For that we use $O$ to denominate the order of magnitude, and $T_i$ to denominate the $ith$ term in the equation, with $i$ increasing from the left to the right. We start by noting that the non-magnetic solution varies typically in a scale of $H$, i.e., $O(\frac{d {\xi}_{si}}{d z})\approx H^{-1}{\xi}_{si}$. For frequencies not too far from $\omega_c$, as those seen in roAp stars, this means that $O(\frac{d {\xi}_{si}}{d z})\approx O((\omega/c) {\xi}_{si}) = O(\omega\sqrt{\rho_0/(\gamma p_0)}{\xi}_{si})$. 

%we find $O(T_1)$ is never smaller than $O((\frac{d p}{pdz} a)^2 \xi_{\bot s}))$ where $a$ is typically $1/2$. In practice $O(T_1)=max[O((\frac{dp}{pdz} a)^2\xi_{\bot s}));\frac{\omega^2 \rho_0 \left|\bm{B}_0\right|^2}{\gamma p_0 B_z^2} \xi_{\bot s}]$. 

Having this in mind we find that for equation (\ref{eq: 2.19}) the order of magnitude of the terms is as follows,

\begin{eqnarray}
O(T_1)\approx O\left(\frac{\omega^2 \rho_0}{\gamma p_0}{\xi}_{\bot s}\right), 
\end{eqnarray}

\begin{eqnarray}
O(T_2)\approx 0\left(\tilde{\beta}\frac{\omega^2 \rho_0}{\gamma p_0}{\xi}_{\bot s}\right), 
\end{eqnarray}

In the magnetically dominated region, $\tilde{\beta}<<1$ and, consequently, $O(T_1)>>O(T_2)$. Thus, equation (\ref{eq: 2.19}) leads, after separating the time and considering the slow component of the displacement, approximately to equation (\ref{eq: 2.25}). Looking at the latter we can conclude, under the assumptions previously made, that
\begin{equation}
\label{eq: 2.23ca}
{\xi}_{\bot s} \approx O(\tilde{\beta}{\xi}_{||s}).
\end{equation}
Therefore, we find that 
\begin{equation}
{\xi}_{\bot s}<<{\xi}_{|| s}\nonumber,
\end{equation}
which was to be expected, since in a magnetically dominated region the slow displacement is almost acoustic and, hence, takes place essentially along the magnetic field lines. 

Combining equation~(\ref{eq: 2.25}) with equation~(\ref{eq: 2.18}), after separating the time and considering the slow component alone,  and re-organizing the terms, we find,
\begin{equation}
\label{eq: 2.34_tese}
\frac{d^2 {\xi}_{|| s} }{d z^2}+\frac{d p_0}{p_0 d z}\frac{d {\xi}_{|| s}}{d z}+\frac{\omega^2 \rho_0}{\gamma p_0}\frac{\left|\bm{B}_0\right|^2}{B_z^2}{\xi}_{||s}=-\frac{d p_0}{p_0 d z}\frac{B_x}{B_z}\frac{d {\xi}_{\bot s}}{d z}.
\end{equation}

Next we determine the order of magnitude of the terms in equation~(\ref{eq: 2.34_tese}). Also here we start by considering the case of $B_x\approx B_z$. Taking into account equation (\ref{eq: 2.23ca}) we can show that $O(T_3)$ and $O(T_4)$ in equation~(\ref{eq: 2.34_tese}) are as follows,

\begin{eqnarray}
O(T_3) \approx O\left(\frac{\omega^2 \rho_0}{\gamma p_0}\xi_{||s}\right),\nonumber
\end{eqnarray}

\begin{eqnarray}
O(T_4) \approx O\left(\frac{1}{p_0}\frac{dp_0}{dz}\sqrt{\frac{\omega^2 \rho_0}{\gamma p_0}}\tilde{\beta}\xi_{|| s}\right) \approx O\left(\frac{\omega^2 \rho_0}{\gamma p_0}\tilde{\beta}\xi_{||s}\right).\nonumber
\end{eqnarray}
Thus, equation~(\ref{eq: 2.34_tese}) can be written as equation (\ref{eq: 2.26}).

The order of magnitude analysis presented above assumed that $B_x\approx B_z$. However, using the solution for $\xi_{||s}$ found in this way (cf.\ equation~(\ref{eq: 2.30c})), and returning to equations (\ref{eq: 2.18}) and (\ref{eq: 2.19}), we can show that this solution is also appropriate in the cases of $B_x$ tending to zero and $B_z$ tending to zero. The former is equivalent to considering the case in the absence of the magnetic field, so it is necessarily in accordance with the analysis made previously. Let us now consider the case of $B_z$ tending to zero. In this case the last term on the right hand side of equation (\ref{eq: 2.29}) dominates and, therefore, $k^2_{||} \approx \frac{\omega^2\rho_0 |\bm{B}_0|^2}{\gamma p_0 B_z^2}$. Going back to the analysis of the order of magnitude, in equation (\ref{eq: 2.19}) we have,   
\begin{eqnarray}
O(T_1)\approx O\left(\frac{\omega^2 \rho_0}{\gamma p_0}\frac{|\bm{B}_0|^2}{B^2_z}{\xi}_{\bot s}\right), 
\end{eqnarray}
\begin{eqnarray}
O(T_2)\approx O\left(\tilde{\beta}\frac{\omega^2 \rho_0}{\gamma p_0}{\xi}_{\bot s}\right). 
\end{eqnarray}
Thus, we can conclude, as in the previous case, that $O(T_1)$ is always much larger than $O(T_2)$, leading back to equation (\ref{eq: 2.25}).
Moreover, from the latter we can conclude that in this case,
\begin{equation}
\label{eq: 2.23c}
{\xi}_{\bot s} \approx O\left(\tilde{\beta}\frac{B_x B_z}{|\bm{B}_0|^2}{\xi}_{||s}\right).
\end{equation}
Therefore, we find, as before,
\begin{equation}
{\xi}_{\bot s}<<{\xi}_{|| s}\nonumber.
\end{equation}

Let us now compare $O(T_3)$ with $O(T_4)$ in equation (\ref{eq: 2.34_tese}) when $B_z$ tends to zero,
\begin{eqnarray}
O(T_3) \approx O\left(\frac{\omega^2 \rho_0}{\gamma p_0}\frac{|\bm{B}_0|^2}{B^2_z}\xi_{||s}\right),\nonumber
\end{eqnarray}
\begin{eqnarray}
O(T_4) \approx O\left(\frac{\omega^2 \rho_0}{\gamma p_0}\tilde{\beta}\frac{B_x^2}{|B_0|^2}\xi_{||s}\right)\nonumber. 
\end{eqnarray}
Therefore, also in this case, $O(T_3)$ is larger than $O(T_4)$, and we can neglect the fourth term in equation (\ref{eq: 2.34_tese}), which leads us back to equation (\ref{eq: 2.26}).
So, we may conclude that the approximate solution expressed by equation (\ref{eq: 2.30c}), derived when considering $B_x \approx B_z$, is in fact an approximate solution for $\xi_{||s}$ in the more general case, since the terms that were neglected in the first case remain small, for this solution, when the condition $B_x \approx B_z$ is no longer satisfied. 

\vspace{0.3cm}

\underline{\textbf{Fast Magnetic Component}}

\vspace{0.3cm}

To derive equations~(\ref{eq: 2.35}) and (\ref{eq: 2.37}) we again consider equation~(\ref{eq: 2.18}), after separating the time and the fast and slow components of the displacement. For that we shall keep in mind the condition $({\bar{\xi}}_{si})^{-1} d{\bar{\xi}}_{si}/dz >> ({\bar{\xi}}_{fi})^{-1} d{\bar{\xi}}_{fi}/dz$ that was used to distinguish the above mentioned components. Since $O(({\xi}_{si})^{-1} d{{\xi}}_{si}/dz) \approx O(\omega\sqrt{\rho_0/\gamma p_0})$, for equation (\ref{eq: 2.18}) we find,
\begin{eqnarray}
O(T_1) < < O(T_5), \nonumber
\end{eqnarray}
\begin{eqnarray}
O(T_3) < < O(T_5). \nonumber
\end{eqnarray}
Thus, equation (\ref{eq: 2.18}) leads, for the fast component of the displacement, approximately to equation (\ref{eq: 2.35}). In addition, comparing the terms in the latter equation we find that, in all cases, the fast component of the displacement perpendicular to the magnetic field is much larger than the fast component of the displacement parallel to the magnetic field, i.e.,
\begin{eqnarray}
\xi_{\bot f}>>\xi_{|| f}\nonumber.
\end{eqnarray}

Therefore, we can say that the fast displacement is almost perpendicular to the direction of the magnetic field. This is not surprising, since in the magnetically dominated region this component of the displacement is expected to be essentially a magnetic wave, only slightly modified by the gas pressure.

Finally, neglecting $\xi_{|| f}$, when compared to $\xi_{\bot f}$, we find that equation~(\ref{eq: 2.19}) leads, for the fast component, to equation~(\ref{eq: 2.37}).

\end{document}